\documentclass[12pt]{article}

\usepackage{epsfig,cite}

\catcode`@=11

\begin{document}



\input paperdef 


\thispagestyle{empty}
\setcounter{page}{0}
\def\thefootnote{\fnsymbol{footnote}}

\begin{flushright}
ANL--HEP--PR--05--109 \hfill
CERN--PH--TH/2005--158\\
DCPT/05/128 \hfill
EFI--05--19\\
FERMILAB--Pub--05-370--T \hfill
IPPP/05/64\\
hep-ph/0511023 \\
\end{flushright}

\mbox{}\vspace{0em}

\begin{center}

{\large\sc {\bf MSSM Higgs Boson Searches at the Tevatron and the LHC:}}

\vspace*{0.3cm}

{\large\sc {\bf Impact of Different Benchmark Scenarios}}

\vspace{0.5cm}

{\sc M.~Carena$^{\,1}$%
\footnote{
email: carena@fnal.gov
}%
, S.~Heinemeyer$^{\,2}$%
\footnote{
email: Sven.Heinemeyer@cern.ch
}%
, C.E.M.~Wagner$^{\,3,4}$%
\footnote{
email: cwagner@hep.anl.gov
}%
~and G.~Weiglein$^{\,5}$%
\footnote{
email: Georg.Weiglein@durham.ac.uk
}%
}

\vspace*{0.8cm}

$^1$ Theoretical Physics Dept., Fermilab,
Batavia, IL 60510-0500, USA

\vspace*{0.3cm}

$^2$ CERN TH Division, Dept.\ of Physics,
CH-1211 Geneva 23, Switzerland

\vspace*{0.3cm}

$^3$ HEP Division, Argonne Natl.\ Lab., 9700 Cass Ave.,
Argonne, IL 60439, USA

\vspace*{0.3cm}

$^4$ 
Enrico Fermi Institute, Univ.\ of Chicago, 5640 Ellis Ave.,
Chicago, IL 60637, USA

\vspace*{0.3cm}

$^5$ IPPP, University of Durham, Durham DH1~3LE, UK

\end{center}

\vspace*{0.5cm}

\begin{abstract}

The Higgs boson search has shifted from LEP2 to the Tevatron and will
subsequently move to the LHC. 
The current limits from the Tevatron and the prospective sensitivities
at the LHC are often interpreted in specific MSSM scenarios. 
For heavy Higgs boson production and subsequent decay into $b \bar b$
or $\tau^+ \tau^-$, the present Tevatron data allow to set limits in the
$\MA$--$\tb$ plane for small $\MA$ and large $\tb$ values. 
Similar channels have been explored for the LHC, where the discovery reach
extends to higher values of $\MA$ and smaller $\tb$.
Searches for MSSM charged Higgs bosons, produced in top decays
or in association with top quarks, have also been investigated at the Tevatron
and the LHC.
We analyze the current Tevatron limits and prospective LHC
sensitivities. We discuss how robust they are with respect to 
variations of the other MSSM parameters and possible 
improvements of the theoretical predictions for Higgs boson production
and decay. It is shown that
the inclusion of supersymmetric radiative corrections
to the production cross sections and decay widths leads to important 
modifications of the present limits on the MSSM parameter space. The
impact on the region where only the lightest MSSM Higgs boson can be
detected at the LHC is also analyzed.
We propose to extend the existing 
benchmark scenarios by including additional
values of the higgsino mass parameter $\mu$. This affects only slightly
the search channels for a SM-like Higgs boson, while having a major 
impact on the searches for non-standard MSSM Higgs bosons.

\end{abstract}

\def\thefootnote{\arabic{footnote}}
\setcounter{footnote}{0}

\newpage


\section{Introduction}

Disentangling the mechanism that controls electroweak symmetry
breaking is one 
of the main tasks of the current and next generation of colliders. 
Among
the most studied candidates in the literature 
are the Higgs mechanism within the Standard Model (SM) or
within the Minimal Supersymmetric Standard Model (MSSM). Contrary to
the SM, two Higgs doublets are required in the MSSM, resulting in five
physical Higgs boson degrees of freedom. 
In the absence of 
explicit $\cp$-violation in the soft supersymmetry-breaking terms
these are the light and heavy $\cp$-even Higgs bosons, $h$ and $H$, the
$\cp$-odd Higgs boson, $A$, and the charged Higgs boson, $H^\pm$.
The Higgs sector of the MSSM can be specified at lowest
order in terms of $\MZ$, $\MA$, and $\tb \equiv v_2/v_1$, the ratio of the
two Higgs vacuum expectation values. 
The masses of the $\cp$-even neutral Higgs bosons and the
charged Higgs boson can be calculated,
including higher-order corrections, 
in terms of the other MSSM parameters. 

After the termination of LEP in the year 2000 (the close-to-final LEP
results can 
be found in \citeres{LEPHiggsSM,LEPHiggsMSSM}), the Higgs boson search
has shifted to the Tevatron and will later be continued at the LHC. 
Due to the large number of 
free parameters, a complete scan of the MSSM parameter space is too
involved. Therefore  the search results at LEP have been
interpreted~\cite{LEPHiggsMSSM} in 
several benchmark scenarios~\cite{benchmark,benchmark2}. Current
analyses at the Tevatron and investigations of the LHC potential also
have been performed in the scenarios proposed in
\citeres{benchmark,benchmark2}. 
The $\mhmax$~scenario
has been used to obtain conservative bounds on
$\tb$ for fixed values of the top-quark mass and the scale of the
supersymmetric particles~\cite{tbexcl}. Besides the
$\mhmax$~scenario and the no-mixing scenario, where a vanishing mixing
in the stop sector is assumed, the
``small~$\aeff$''~scenario and the ``gluophobic Higgs scenario'' have
been investigated~\cite{schumi}. While the latter one exhibits a
strong suppression 
of the $ggh$ coupling over large parts of the $\MA$--$\tb$ parameter
space, the small~$\aeff$~scenario has strongly reduced couplings of
the light $\cp$-even Higgs boson to bottom-type fermions up to 
$\MA \lsim 350 \gev$.
These scenarios are conceived to study particular cases of 
challenging and interesting phenomenology
in the searches for the SM-like Higgs boson, i.e.\ 
mostly the light $\cp$-even Higgs boson. 

The current searches at the Tevatron are not yet sensitive to a 
SM-like Higgs in the mass region allowed by the LEP
exclusion bounds~\cite{LEPHiggsSM,LEPHiggsMSSM}. On the other hand,
scenarios with enhanced Higgs boson production cross sections can be
probed already with the currently accumulated luminosity. Enhanced
production cross sections can occur in 
particular for low $\MA$ in combination with large $\tb$ due to the
enhanced couplings of the Higgs bosons to down-type fermions. 
The corresponding limits on the Higgs production cross section times
branching ratio of the Higgs decay into down-type fermions can be
interpreted in MSSM benchmark scenarios. Limits
from Run~II of the Tevatron have recently been published for the
following channels~\cite{D0bounds,CDFbounds,Tevcharged} (here and in the
following $\phi$ denotes all three neutral MSSM Higgs bosons, 
$\phi = h, H, A$):
\BEA
\label{scen1}
    && b \bar b \phi,  \phi \to b \bar b 
       ~(\mbox{with one additional tagged } b \mbox{ jet}) , \\[.3em] 
\label{scen2}
       && p \bar p \to \phi \to \tau^+\tau^-  
        ~(\mbox{inclusive}) , \\[.3em]
\label{scen3}
     &&p \bar p \to t \bar t \to H^\pm W^\mp \, b \bar b,
     H^{\pm} \to \tau \nu_{\tau} ~.
\EEA
The obtained cross section limits have been interpreted in the
$\mhmax$ and the no-mixing scenario with a value for the higgsino mass
parameter of $\mu = - 200 \gev$~\cite{D0bounds} and  
$\mu = \pm 200 \gev$~\cite{CDFbounds}.
In these scenarios for $\MA \approx 100 \gev$ the limits on $\tb$ are
$\tb \lsim 50$.

In this article, we investigate the dependence of the CDF and D0
exclusion bounds in the $\MA$--$\tb$ plane on the parameters entering
through the most relevant supersymmetric radiative corrections
in the theoretical predictions for Higgs boson
production and decay processes.
We will show that the bounds obtained from the
$b \bar b \phi, \phi \to b \bar b$ channel depend
very sensitively on the radiative corrections affecting the relation
between the bottom quark mass and the bottom Yukawa coupling.%
\footnote{
We concentrate here on the effects of supersymmetric radiative
corrections. For a recent account of uncertainties related to parton
distribution functions, see e.g.~\citere{Belyaev:2005nu}.
}%
~In the channels with $\tau^+ \tau^-$ final states, on the other hand,
compensations between large corrections in the Higgs production and
the Higgs decay occur. 
In this context we investigate the impact of a
large radiative correction in the $gg \to \phi$ 
production process that had previously been omitted.
 
In order to reflect the impact of the corrections to the bottom Yukawa
coupling on the exclusion bounds we suggest to supplement the existing
$\mhmax$~ and no-mixing scenarios, mostly designed to
search for the light $\cp$-even MSSM Higgs boson, $h$, with additional
values for the higgsino mass parameter $\mu$. In fact, varying the
value and sign of $\mu$, while keeping fixed the values of the gluino
mass and the common third generation squark mass parameter $\msusy$,
demonstrates 
the effect of the radiative corrections on the production and
decay processes.  
The scenarios discussed here are designed specifically to
study the MSSM Higgs sector without assuming any particular soft
supersymmetry-breaking scenario and taking into account constraints 
only from the Higgs boson sector itself. In particular, constraints
from requiring the correct cold dark matter density, 
$\br(b \to s \ga)$ or $(g - 2)_\mu$, 
which depend on other parameters of the theory, are not crucial in defining
the Higgs
boson sector, and may be avoided. However, we also include a brief discussion
of the ``constrained-$\mhmax$'' scenario, which in the case of minimal
flavor violation and positive values of $\mu$ leads to a better
agreement with the constraints from $\br(b \to s \ga)$.

We also study the non-standard MSSM Higgs boson search sensitivity 
at the LHC, focusing on the processes 
$pp \to H/A +X, \, H/A \to \tau^+ \tau^-$ and 
$pp \to t H^{\pm} +X, \, H^{\pm} \to \tau \nu_{\tau}$,
and stress the relevance of the proper inclusion of supersymmetric
radiative corrections to the production cross sections and decay widths.
We show the impact of these corrections by investigating the variation
of the Higgs boson discovery reach in the benchmark scenarios for
different values of $\mu$. 
In particular, we discuss the resulting
modification of the parameter region in which only the
light $\cp$-even  MSSM Higgs boson can be detected at the LHC.

The paper is organized as follows: Section~2 gives
a summary of the most relevant supersymmetric
radiative corrections to the Higgs boson production cross
section and decay widths, while also introducing our notation.
In section~3 we analyze the impact of these radiative corrections on
the current Tevatron limits in the large $\tb$ region, as well as on 
the future LHC reach for the heavy, 
non-standard, MSSM Higgs bosons. 
In section 4, based on the results of section~3, we propose an
extension of the existing benchmark scenarios. The conclusions are
presented in section~5.


\section{Predictions for Higgs boson production and decay processes}

\subsection{Notation and renormalization}

The tree-level values for the $\cp$-even Higgs bosons of the
MSSM, $\mh$ and $\mH$, are determined by $\tb$, 
the $\cp$-odd Higgs-boson mass $\MA$, and the $Z$ boson mass $\MZ$. 
The mass of the charged Higgs boson, $\MHp$, is given in terms of $\MA$
and the $W$ boson mass, $\MW$.
Beyond the tree-level, the main correction to the Higgs boson masses
stems from the 
$t/\Stop$ sector, and for large values of $\tb$ also from the 
$b/\Sbot$ sector.

In order to fix our notations, we list the conventions for the inputs
from the scalar top and scalar bottom sector of the MSSM:
the mass matrices in the basis of the current eigenstates $\StopL, \StopR$ and
$\SbotL, \SbotR$ are given by
\BEA
\label{stopmassmatrix}
{\cal M}^2_{\Stop} &=&
  \ML \MstL^2 + \mt^2 + \CZb (\edz - \frac{2}{3} \sw^2) \MZ^2 &
      \mt \Xt \\
      \mt \Xt &
      \MstR^2 + \mt^2 + \frac{2}{3} \CZb \sw^2 \MZ^2 
  \MR, \\
&& \non \\
\label{sbotmassmatrix}
{\cal M}^2_{\Sbot} &=&
  \ML \MsbL^2 + \mb^2 + \CZb (-\edz + \frac{1}{3} \sw^2) \MZ^2 &
      \mb \Xb \\
      \mb \Xb &
      \MsbR^2 + \mb^2 - \frac{1}{3} \CZb \sw^2 \MZ^2 
  \MR,
\EEA
where 
\BE
\mt \Xt = \mt (\At - \mu \CTb) , \quad
\mb\, \Xb = \mb\, (\Ab - \mu \Tb) .
\label{eq:mtlr}
\EE
Here $\At$ denotes the trilinear Higgs--stop coupling, $\Ab$ denotes the
Higgs--sbottom coupling, and $\mu$ is the higgsino mass parameter.

SU(2) gauge invariance leads to the relation
\BE
\MstL = \MsbL .
\EE
For the numerical evaluation, a convenient choice is
\BE
\MstL = \MsbL = \MstR = \MsbR =: \msusy .
\label{eq:msusy}
\EE
Concerning analyses for the case where $\MstR \neq \MstL \neq \MsbR$, see e.g.\
\citeres{stefanCM,mhiggslong}. It has been shown that the upper bound on $\mh$ 
obtained using \refeq{eq:msusy} is the same as for the more general
case, provided that $\msusy$ is identified with the heaviest mass of
$\MstR, \MstL, \MsbR$~\cite{mhiggslong}. 
We furthermore use the short-hand notation
\BE
\ms^2 := \msusy^2 + \mt^2~.
\EE

Accordingly, the most important parameters for the corrections in the
Higgs sector
are $\mt$, $\msusy$, $\Xt$, and $\Xb$. The Higgs sector observables
furthermore depend on the SU(2) gaugino mass
parameter, $M_2$. The other gaugino mass parameter, $M_1$, is usually
fixed via the GUT relation 
\BE
M_1 = \frac{5}{3} \frac{\sw^2}{\cw^2} M_2.
\EE
At the two-loop level also the gluino mass, $\mgl$, enters the
predictions for the Higgs-boson masses.

\bigskip
Corrections to the MSSM Higgs boson sector have been evaluated in
several approaches. 
The status of the available corrections to the masses and mixing
angles in the MSSM Higgs sector (with real parameters) 
can be summarized as follows. For the
one-loop part, the complete result within the MSSM is 
known~\cite{ERZ,mhiggsf1lA,mhiggsf1lB,mhiggsf1lC}. The by far dominant
one-loop contribution is the \order{\alt} term due to top and stop 
loops ($\alt \equiv h_t^2 / (4 \pi)$, $h_t$ being the 
top-quark Yukawa coupling).
Concerning the two-loop
effects, their computation is quite advanced and has now reached a
stage such that all the presumably dominant contributions are 
known~\cite{mhiggsEP1b,mhiggsRG1a,mhiggsRG1,HHH,mhiggsletter,mhiggslong,mhiggsEP0,mhiggsEP1,mhiggsEP3,mhiggsEP2,mhiggsEP4,mhiggsFD2,deltamb1,deltamb2,deltamb2b,mhiggsEP4b}.
They include (evaluated for vanishing external momenta) the strong
corrections, usually indicated as \order{\alt\als}, and Yukawa
corrections, \order{\alt^2}, to the dominant one-loop \order{\alt}
term, as well as the strong corrections to the bottom/sbottom one-loop
\order{\alb} term ($\alb \equiv h_b^2 / (4\pi)$), i.e.\ the
\order{\alb\als} contribution. The latter can be relevant for large
values of $\tb$. 
For the (s)bottom corrections the all-order resummation of the
$\tb$-enhanced terms, \order{\alb(\als\tb)^n}, 
has also been computed. Recently the \order{\alt\alb} and
\order{\alb^2} corrections have been obtained. 
The remaining theoretical uncertainty on the light $\cp$-even Higgs
boson mass has been estimated to be below 
$\sim 3 \gev$~\cite{mhiggsAEC,PomssmRep}. 
The above calculations have been implemented into public 
codes. The program 
{\tt FeynHiggs}~\cite{feynhiggs,feynhiggs1.2,feynhiggs2,feynhiggs2.3} 
is based on the results obtained in the Feynman-diagrammatic (FD)
approach~\cite{mhiggsletter,mhiggslong,mhiggsAEC,mhiggsEP4b}. It
includes all the above corrections. The code 
{\tt CPsuperH}~\cite{cpsh} is based on the renormalization group (RG)
improved effective potential
approach~\cite{mhiggsRG1a,mhiggsRG1,bse}. 
For the MSSM with real parameters the two codes can differ by up to
$\sim 4 \gev$ for the light $\cp$-even Higgs boson mass, mostly due to
formally subleading two-loop corrections that are included only in 
{\tt FeynHiggs}. For the MSSM with complex parameters the phase
dependence at the two-loop level is included in a more advanced
way~\cite{mhiggsCPXRG} in {\tt CPsuperH}, but, on the other hand, 
{\tt CPsuperH} does not contain all the subleading one-loop contributions
that are included~\cite{mhiggsCPXFD,habilSH} in {\tt FeynHiggs}. 
Most recently a full two-loop effective potential 
calculation (including even the momentum dependence for the leading
pieces) has been published~\cite{mhiggsEP5}. However, no computer code
is publicly available. In the following we will concentrate on the MSSM
with real parameters.

\medskip
It should be noted in this context that the FD result has been obtained
in the on-shell (OS) renormalization scheme, whereas the RG result has been 
calculated using the \msbar\ scheme; see \citeres{bse,mhiggslle} for a
detailed comparison. Owing to the different
schemes used in the FD and the RG approach for the
renormalization in the scalar top sector, the
parameters $\Xt$ and $\msusy$ are also scheme-dependent
in the two approaches. This difference between the corresponding
parameters has to be taken into account 
when comparing the results of the two approaches. In a simple approximation
the relation between the parameters in the different schemes is at
\order{\als} given by~\cite{bse}
\BEA
\ms^{2, \MS} &\approx& \ms^{2, \OS} 
 - \frac{8}{3} \frac{\al_s}{\pi} \ms^2 , 
\label{eq:msms} \\
\Xt^{\MS} &\approx& X_t^{\OS} + \frac{\al_s}{3 \pi} \ms 
   \left(8 + 4 \frac{X_t}{\ms} - 3 \frac{X_t}{\ms} 
   \log\left(\frac{\mt^2}{\ms^2}\right) \right),
\label{eq:xtms} 
\EEA 
where in the terms proportional to $\al_s$ it is not necessary to
distinguish between \msbar\ and on-shell quantities, since the
difference is of higher order.
The \msbar\ top-quark mass, $\mt^{\MS}(m_t) \equiv \mtms$, 
is related to the top-quark pole
mass, $\mt^{\OS} \equiv \mt$, in \order{\als} by
\BE
\mtms = \frac{\mt}{1 + \frac{4}{3\,\pi} \als(\mt)}~.
\label{mtrun}
\EE
While the resulting shift in the parameter $\msusy$ turns out to be 
relatively small in general, sizable differences can occur between the 
numerical values of 
$\Xt$ in the two schemes, see \citeres{mhiggslong,bse}. For this reason
we specify below different values for $\Xt$ within the two approaches.


\subsection{Leading effects from the bottom/sbottom sector}

The relation between the bottom-quark mass and the Yukawa coupling
$h_b$, which controls also the interaction between the Higgs fields and
the sbottom quarks, reads at lowest order $\mb =h_b v_1$. 
This relation is affected at \onel\ order by large radiative
corrections \cite{deltamb1,deltamb2,deltamb2b,deltamb3},
proportional to $h_b v_2$,  
in general giving rise to $\tb$-enhanced contributions.
These terms proportional to $v_2$, often called threshold
corrections to the bottom mass, are generated either by
gluino--sbottom \onel\ diagrams (resulting in \order{\alb\als}
corrections to the Higgs masses), or by chargino--stop loops (giving
\order{\alb\alt} corrections). Because the $\tb$-enhanced
contributions can be numerically relevant, an accurate determination
of $h_b$ from the experimental value of the bottom mass requires a
resummation of such effects to all orders in the perturbative
expansion, as described in \citeres{deltamb2,deltamb2b}.

The leading effects are included in the effective Lagrangian
formalism developed in \citere{deltamb2}.
Numerically this is by far the dominant part of the
contributions from the sbottom sector (see also
\citeres{mhiggsEP4,mhiggsEP4b,mhiggsFD2}). The dominant contributions
arise from the loop-induced coupling of $H_u$ (the Higgs field that
couples at the tree-level to up-type fermions only) to the down-type
fermions. The effective Lagrangian is given by
\BEA
\cL = \frac{g}{2\MW} \frac{\mbms}{1 + \db} \Bigg[ 
&& \tb\; A \, i \, \bar b \ga_5 b 
   + \wz \, V_{tb} \, \tb \; H^+ \bar{t}_L b_R \non \\
&+& \KL \frac{\Sa}{\Cb} - \db \frac{\Ca}{\Sbe} \KR h \bar{b}_L b_R 
                                                               \non \\
&-& \KL \frac{\Ca}{\Cb} + \db \frac{\Sa}{\Sbe} \KR H \bar{b}_L b_R
    \Bigg] + {\rm h.c.}~.
\label{effL}
\EEA
Here $\mbms$ denotes the running bottom quark mass including SM QCD
corrections. In the numerical evaluations obtained with 
{\tt FeynHiggs} below we choose  
$\mbms = \mbms(\mt) \approx 2.97 \gev$. 
The prefactor $1/(1 + \db)$ in \refeq{effL} arises from the
resummation of the leading corrections to all orders. 
The additional terms $\sim \db$ in the $h\bar b b$ and $H\bar b b$
couplings arise from the mixing and coupling of the ``other'' Higgs
boson, $H$ and $h$, respectively, to the $b$~quarks.

As explained above, 
the function $\db$ consists of two main contributions, 
an \order{\als} correction from a
sbottom--gluino loop and an \order{\alt} correction
from a stop--higgsino loop. The explicit
form of $\db$ in the limit of $M_S \gg \mt$ and $\tb \gg 1$
reads~\cite{deltamb1}
\BE
\db = \frac{2\als}{3\,\pi} \, \mgl \, \mu \, \tb \,
                    \times \, I(\msbe, \msbz, \mgl) +
      \frac{\alt}{4\,\pi} \, \At \, \mu \, \tb \,
                    \times \, I(\mste, \mstz, \mu) ~.
\label{def:dmb}
\end{equation}
The function $I$ is given by
\BEA
I(a, b, c) &=& \ed{(a^2 - b^2)(b^2 - c^2)(a^2 - c^2)} \,
               \KL a^2 b^2 \log\frac{a^2}{b^2} +
                   b^2 c^2 \log\frac{b^2}{c^2} +
                   c^2 a^2 \log\frac{c^2}{a^2} \KR \\
 &\sim& \ed{\mbox{max}(a^2, b^2, c^2)} ~. \non
\EEA
The large $\Sbot-\gl$~loops are resummed to all orders of
$(\als\tb)^n$ via the inclusion of $\db$~\cite{deltamb1,deltamb2,deltamb2b}.
The leading electroweak contributions are taken into account via the
second term in \refeq{def:dmb}.

For large values of $\tb$ and the ratios of $\mu \mgl/ \msusy^2$ and 
$\mu \At/ \msusy^2$,  the $\db$ correction can
become very important. Considering positve values of $\At$ and $\mgl$, the 
sign of the $\db$ term is governed by the sign of $\mu$. 
Cancellations can occur if $\At$ and $\mgl$ have opposite signs.
For $\mu, \mgl, \At > 0$ the $\db$ correction is positive, leading 
to a suppression of the bottom Yukawa coupling. On the other hand,
for negative values of 
$\db$, the bottom Yukawa coupling
may be strongly enhanced and can even acquire 
non-perturbative values when $\db \to -1$.


\subsection{Impact on Higgs production and decay at large $\tb$}

\label{sec:impact}

Higgs-boson production and decay processes at the Tevatron and the LHC
can be affected by different kinds of large radiative corrections. The
SM and MSSM corrections to the production channel $gg \to \phi$
have been calculated in \citeres{ggHSM,ggHMSSM}, SM corrections to the 
$b\bar b\phi$ channel have been evaluated 
in \citeres{D0bbhSM,bbHSM,robikilgore}. Higgs decays to $b \bar b$ and to 
$\tau^+\tau^-$ within the SM and MSSM have been evaluated including
higher-order corrections in \citeres{hff,mhiggsf1lC,habilSH}.
Besides the process-specific corrections 
to the production and decay processes, 
large Higgs-boson propagator contributions have an impact on
the Higgs-boson couplings. 
For large $\tb$ the supersymmetric radiative corrections to the
bottom Yukawa coupling described above become particularly 
important~\cite{deltamb4}. 
Their main effect on the Higgs-boson production and decay processes  
can be understood from the way the 
leading contribution $\db$ enters. In the following we present simple 
analytic approximation formulae for the most relevant Higgs-boson 
production and decay processes. They are meant for illustration only so
that the impact of the $\db$ corrections can easily be traced
(for a discussion of possible enhancement factors for MSSM Higgs-boson
production processes at the Tevatron and the LHC, see also 
\citeres{SolaChargedHiggs,Belyaev:2005ct}).
In our 
numerical analysis below, we use the full result from {\tt FeynHiggs} rather 
than the simple formulae presented in this section. 
No relevant modification
to these results would be obtained using {\tt CPsuperH}.

\bigskip
We begin with a simple approximate formula that represents well the MSSM
parametric variation of the decay rate of the $\cp$-odd Higgs boson
in the large $\tb$ regime. One should recall, for that purpose,
that in this regime the $\cp$-odd Higgs boson decays mainly into 
$\tau$-leptons and bottom-quarks, and that the partial decay widths are
proportional to the square of the Yukawa couplings evaluated at
an energy scale of about the Higgs boson mass. Moreover, for Higgs
boson masses of the order of 100~GeV, the approximate relations
$m_b(\MA)^2 \simeq 9$~GeV$^2$, and $m_\tau(\MA)^2 \simeq 3$~GeV$^2$
hold.  Hence, since the number of colors is $N_c = 3$, 
for heavy supersymmetric particles, with masses far above
the Higgs boson mass scale, one has 
\begin{eqnarray}
\label{eq:BRAbb}
{\rm BR}(A \to b \bar{b}) & \simeq  & 
\frac{ 9}{\left(1 + \db \right)^2 + 9} ~ , \\
{\rm BR}(A \to \tau^+\tau^-) & \simeq &
\frac{\left(1 + \db\right)^2}{\left(1 + \db \right)^2 + 9} ~ .
\end{eqnarray}

On the other hand, the production cross section for a $\cp$-odd Higgs boson
produced in association with a pair of bottom quarks is proportional
to the square of the bottom Yukawa coupling and therefore is proportional
to $\tan^2\be/(1 + \db)^2$. Also in the gluon fusion channel,
the dominant contribution in the large $\tb$ regime
is governed by the bottom quark loops, and therefore is also proportional
to the square of the bottom Yukawa coupling. Since the top-quark coupling is
suppressed by either loop-corrections or inverse powers of $\tb$,
the leading top-quark correction arises from interference terms between the 
top-quark
and bottom-quark loop diagrams. We have checked that these interference
terms lead to corrections smaller than one percent (a few percent) for
values of $\tb \simeq 50$ (20). These corrections are small, 
of the order of other subleading corrections not included in our analysis, 
and lead to a very small modification of the current Tevatron limits
(a small shift, smaller than $\De\tb \sim 1$, in the limit on $\tb$). 
They have been neglected in the 
CDF analysis of the 
$\sigma(p\bar{p} \to \phi) \times {\rm BR}(\phi \to \tau^+ \tau^-)$
process. We shall omit these corrections in 
the analytical formulae presented in this section 
and also in the numerical analysis below. However, including them would
be straightforward (leading, as discussed above, to a very small
modification of the allowed parameter space). Hence, the total 
production rate of bottom quarks and $\tau$ pairs
mediated by the production of a $\cp$-odd Higgs boson in the large $\tb$
regime is approximately given by
\begin{eqnarray}
\label{eq:bbA}
\sigma(b\bar{b} A) \times {\rm BR}(A \to b \bar{b}) &\simeq&
\sigma(b\bar{b} A)_{\rm SM} \;
\frac{\tan^2\be}{\left(1 + \db \right)^2} \times
\frac{ 9}{
\left(1 + \db \right)^2 + 9} ~, \\
\label{eq:Atautau}
\sigma(gg, b\bar{b} \to A) \times {\rm BR}(A \to \tau^+ \tau^-) &\simeq&
\sigma(gg, b\bar{b} \to A)_{\rm SM} \;
\frac{\tan^2\be}{
\left(1 + \db \right)^2 + 9} ~,
\end{eqnarray} 
where $\sigma(b\bar{b}A)_{\rm SM}$ and $\sigma(gg, b\bar{b} \to A)_{\rm SM}$ 
denote the values of the corresponding SM Higgs boson production cross
sections for a Higgs boson mass equal to $\MA$.

As a consequence, the $b\bar{b}$ production rate depends sensitively on
$\db$ because of the factor $1/(1 + \db)^2$, while this leading
dependence on $\db$ cancels out in the $\tau^+\tau^-$ production rate.
There is still a subdominant parametric dependence in the $\tau^+\tau^-$ 
production rate on $\db$ that may lead
to variations of a few tens of percent of the $\tau$-pair production
rate (compared to variations of the rate by up to 
factors of a few in the case of bottom-quark pair
production).

The formulae above apply, within a good approximation, also to the
non-standard $\cp$-even 
Higgs boson in the large $\tb$ regime. Indeed, unless the $\cp$-odd
Higgs mass is within a small regime of masses of about
$\mhmax \simeq$~120--130~GeV,
the mixing of the two $\cp$-even Higgs bosons is small and, for 
$\MA > \mhmax$ ($\MA < \mhmax$), $\cos\al \simeq \sin\be$
($\sin\al \simeq -\sin\be$). In addition, this non-standard Higgs
boson becomes degenerate in mass with the $\cp$-odd Higgs scalar.
Therefore, the production and decay
rates of $H$ ($h$) are governed by similar formulae as the ones presented
above, leading to an approximate enhancement of a factor 2 of the production
rates with respect to the ones that would be obtained in the case of the
single production of the $\cp$-odd Higgs boson as given in
\refeqs{eq:bbA}, (\ref{eq:Atautau}). The same is true in the
region where all three neutral Higgs bosons are approximately
mass-degenerate, $\mh \simeq \mH \simeq \MA$. In this case the combined
contribution of $h$ and $H$ to the production and decay rates
approximately equals the contribution of the $\cp$-odd Higgs boson.

\bigskip
Besides the effects discussed above, 
additional radiative corrections can be important in the search for
non-standard MSSM Higgs bosons.
In particular,
there are radiative corrections to the mass difference
between $\mH$ (the non-SM like $\cp$-even Higgs boson) 
and $\MA$~\cite{deltamb4}. If the two states are roughly
mass-degenerate, one obtains a 
factor of~2 in the production rate, as outlined above.
In the case of small mixing between the two $\cp$-even states, the
mass difference is approximately given by 
\begin{equation}
\mH^2 - \MA^2 \simeq
- \frac{\GF}{4 \, \wz \, \pi^2} \KKL 
  \mt^4 \; \frac{(\mu \; \At)^2}{\msusy^4}
+ \frac{\mb^4}{(1 + \db)^4} \frac{(\mu \; \Ab)^2}{\msusy^4} \KKR~,
\label{eq:deltamHmA}
\end{equation}
where $\GF$ is the Fermi constant, and $\mq$ denotes the running quark
masses. 
For large values of $(\mu \At)$ and/or
large values of $(\mu \Ab)$ and $\tb$, the mass difference
becomes so large that the signals arising from the
production of the $\cp$-even and the $\cp$-odd Higgs bosons can no
longer simply be added, leading to a modification of the Higgs search
sensitivity at the Tevatron and the LHC. A further important set of
corrections are contributions from the sbottom sector
giving rise to a large downward shift in
the mass of the light $\cp$-even Higgs boson,
\begin{equation}
\delta \mh^2 \simeq
-\frac{\GF}{4 \, \wz \, \pi^2} \frac{\mb^4}{(1 + \db)^4}
 \left(\frac{\mu}{\msusy}\right)^4 .
\label{eq:deltamh}
\end{equation}
For large values of $\mu$ and $\tb$ these corrections can shift 
the prediction for $\mh$ below the 
experimental bound from LEP~\cite{LEPHiggsSM,LEPHiggsMSSM}. This can
happen in particular for small 
mixing in the stop sector, for which the LEP bounds exclude a
significant part of the parameter space.
Finally, there are radiative corrections affecting the mixing of the two 
$\cp$-even Higgs states that are not included in the above expressions.
In particular, bottom-Yukawa induced corrections lead
to an enhancement (suppression) of the mixing between the $\cp$-even
Higgs bosons
for large and negative (positive) values of 
$\mb^4/(1 + \db)^4 \; \mu^3 \Ab/\msusy^4$.
An enhancement of the mixing between $h$ and $H$ implies that 
the mass difference between 
$\mH$ and $\MA$ is pushed to larger values than those given in 
\refeq{eq:deltamHmA} and that $\mh$ receives a further downward shift in
addition to the correction in \refeq{eq:deltamh}.

\bigskip
We now turn to the production and decay processes of the charged Higgs
boson. In the MSSM, the masses and couplings of the 
charged Higgs boson in the large $\tb$ regime
are closely related to the ones of the $\cp$-odd Higgs
boson. 
The tree-level relation
$\MHp^2 = \MA^2 + \MW^2$
receives sizable corrections  
for large values of $\tb$, $\mu$, $\At$ and $\Ab$,
\BEA
\label{eq:MHpmHO}
\MA^2 &\simeq& \MHp^2 - \MW^2 \\
&+& \frac{3 \, \GF}{8 \, \wz \, \pi^2}
\KKL
- \frac{\mt^2 \, \mb^2}{(1 + \db)^2}
\left( 4\log\left(\frac{M_S^2}{m_t^2} \right) + 2 A_{tb} \right) 
+ \KL \mt^4 + \frac{\mb^4}{(1 + \db)^4} \KR 
\left(\frac{\mu}{\msusy}\right)^2 \KKR~, \non
\EEA
with~\cite{mhiggsRG1a} 
\begin{equation}
\label{eq:Atb}
A_{tb} =
\frac{1}{6} \left[- \frac{6 \mu^2}{\msusy^2}
- \frac{ \left(\mu^2 - \Ab \At\right)^2}{\msusy^4} +
\frac{3 \left(\At + \Ab \right)^2}{\msusy^2} \right].
\end{equation}
The coupling of the charged Higgs boson to a top and a bottom quark
at large values of $\tb$ is governed by the bottom Yukawa coupling and is
therefore affected by the same $\db$ corrections that
appear in the couplings of
the non-standard neutral MSSM Higgs bosons~\cite{deltamb2}.

The relevant channels for charged Higgs boson searches
depend on its mass.
For values of $\MHp$ smaller than the top-quark mass,
searches at hadron colliders
concentrate on the possible emission of the charged
Higgs boson from top-quark decays. In this case, for
large values of $\tb$, the charged Higgs decays
predominantly into a $\tau$ lepton and a neutrino, i.e.\ one has to a
good approximation
\BE
\br(H^\pm \to \tau \nu_\tau) \approx 1~.
\EE
The partial decay width of the top quark into a charged Higgs
and a bottom quark is proportional to the square of
the bottom Yukawa coupling and therefore scales with
$\tan^2\be/(1 + \db)^2$, see e.g.~\citere{deltamb2}. 

For values of the charged Higgs mass larger than $\mt$,
instead, the most efficient production channel is the
one of a charged Higgs associated with a top quark 
(mediated, for instance, by gluon-bottom fusion). In
this case, the production cross section is proportional
to the square of the bottom-quark Yukawa coupling. The
branching ratio of the charged Higgs decay into a 
$\tau$ lepton and a neutrino is, apart from threshold corrections, 
governed by a similar formula
as the branching ratio of the decay of the $\cp$-odd Higgs
boson into $\tau$-pairs, namely
\begin{equation}
\br(H^{\pm} \to \tau \nu_\tau) \simeq \frac{(1+\db)^2}{(1+\db)^2 + 9
\; (1 - r_t)^2}~,
\label{def:brcharged}
\end{equation}
where the factor $(1 - r_t)^2$ is associated with threshold corrections,
and $r_t = \mt^2/\MHp^2$.

\bigskip
As mentioned above, our
numerical analysis will be based on the complete expressions for the
Higgs couplings rather than on the simple approximation formulae given
in this section.


\section{Interpretation of cross section limits in MSSM\\ scenarios}
\label{sec:interpretation}

\subsection{Limits at the Tevatron}

The D0 and CDF Collaborations have recently published cross section
limits from the Higgs search at the Tevatron in the channel where at
least three bottom quarks are identified in the final state 
($b \bar b \phi, \phi \to b \bar b$)~\cite{D0bounds}
and in the inclusive channel with $\tau^+\tau^-$ final
states ($p \bar p \to \phi \to \tau^+\tau^-$)~\cite{CDFbounds}.
The CDF Collaboration has also done analyses searching for a charged Higgs
boson in top-quark decays~\cite{Tevcharged}. 
For a Higgs boson with a mass of about 120~GeV, the D0 Collaboration
excludes a cross section of about 30~pb in the 
$b \bar b \phi , \phi \to b \bar b$ channel with a luminosity of
260~pb$^{-1}$~\cite{D0bounds}, and the CDF Collaboration excludes a
cross section of about 15~pb in the $p \bar p \to \phi \to \tau^+\tau^-$ 
channel with a luminosity of 310~pb$^{-1}$~\cite{CDFbounds}.
While the cross section for a SM Higgs boson is significantly below the
above limits, a large enhancement of these cross sections is possible in
the MSSM.

It is therefore of interest to interpret the cross section limits within
the MSSM parameter space. Since the Higgs sector of the MSSM is
characterised by two new parameters at lowest order, conventionally chosen
as $\MA$ and $\tb$, one usually displays the limits in the $\MA$--$\tb$
plane (for $\cp$-violating scenarios one normally chooses the
$M_H^{\pm}$--$\tb$ plane). As the whole structure of the MSSM enters via
radiative corrections, the limits in the $\MA$--$\tb$ plane depend on the
other parameters of the model. One usually chooses certain benchmark
scenarios to fix the other MSSM parameters~\cite{benchmark,benchmark2}.
In order to understand the physical meaning of the exclusion bounds in
the $\MA$--$\tb$ plane it is important to investigate how sensitively they
depend on the values of the other MSSM parameters, i.e.\ on the choice
of the benchmark scenarios.


\subsubsection{Limits from the process $b \bar b \phi, \phi \to b \bar b$}

The D0 Collaboration has presented the limits in the $\MA$--$\tb$ plane
obtained from the $b \bar b \phi, \phi \to b \bar b$ channel for the
$\mhmax$ and no-mixing scenarios as defined in \citere{benchmark}.
The $\mhmax$ scenario according to the definition of \citere{benchmark} reads
\BEA
\mt &=& 174.3 \gev, \non \\
\msusy &=& 1000 \gev, \non \\
\mu &=& -200 \gev, \non \\
M_2 &=& 200 \gev, \non \\
\Xt^{\OS} &=& 2\, \msusy  \; \mbox{(FD calculation)}, \non \\
\Xt^{\MS} &=& \sqrt{6}\, \msusy \; \mbox{(RG calculation)} \non \\ 
\Ab &=& \At, \non \\
\mgl &=& 0.8\,\msusy~.
\label{oldmhmax}
\EEA
The no-mixing scenario defined in \citere{benchmark} differs from the
$\mhmax$ scenario only in
\BE
\Xt = 0 \; \mbox{(FD/RG calculation)}~.
\label{oldnomix}
\EE
The condition $\Ab = \At$ implies that the different mixing in the stop
sector gives rise to a difference between the two scenarios also in the
sbottom sector. 
The definition of the $\mhmax$ and no-mixing scenarios given in
\citere{benchmark} was later updated 
in \citere{benchmark2}, see the discussion below.

For their analysis, the D0 Collaboration has used the
following approximate formula~\cite{D0bounds},
\BE
\si(b \bar b \phi) \times {\rm BR}(\phi \to b \bar b) =
2 \; \si(b \bar b \phi)_{\rm SM} \;
\frac{\tan^2\be}{\left(1 + \db \right)^2} \times
\frac{ 9}{
\left(1 + \db \right)^2 + 9}~,
\label{eq:d0formula}
\EE
which follows from \refeq{eq:bbA} and the discussion in
\refse{sec:impact}. The cross section 
$\si(b \bar b \phi)_{\rm SM}$ has been evaluated with the code
of~\citere{D0bbhSM}, while $\db$ has been calculated using 
{\tt CPsuperH}~\cite{cpsh}.
{}From the discussion in \refse{sec:impact} it follows  that the choice
of negative values of $\mu$ 
leads to an enhancement of the bottom
Yukawa coupling and therefore to an enhancement of the signal cross
section in \refeq{eq:d0formula}.
For $\tb = 50$ the quantity $\db$ takes on the following values in
the $\mhmax$ and no-mixing scenarios as defined in
\refeqs{oldmhmax}, (\ref{oldnomix}),
\BEA
\mbox{$\mhmax$ scenario, $\mu = -200 \gev$, $\tb = 50$} &:&
\db = -0.21~, \label{eq:db1} \\
\mbox{no-mixing scen., $\mu = -200 \gev$, $\msusy = 1000 \gev$,
                                                 $\tb = 50$} &:&
\db = -0.10~. \label{eq:db2}
\EEA
While the \order{\als} contribution to $\db$, see \refeq{def:dmb},
is practically the same in the two scenarios, the \order{\alt}
contribution to $\db$ in the $\mhmax$ scenario differs significantly
from the one in the no-mixing scenario. In the $\mhmax$ scenario the 
\order{\alt} contribution to $\db$ is about as large as the \order{\als}
contribution. In the no-mixing scenario, on the other hand, the 
\order{\alt} contribution to $\db$ is very small, because $\At$ is close
to zero in this case.
Reversing the sign of $\mu$ in \refeqs{eq:db1}, (\ref{eq:db2}) 
reverses the sign of $\db$, leading
therefore to a significant suppression of the signal cross section in
\refeq{eq:d0formula} for the same values of the other MSSM parameters.

The predictions for $b \bar b \phi$, $\phi \to b \bar b$ evaluated with
{\tt FeynHiggs} have been compared with the exclusion bound for
$\si \times \br$ as given in \citere{D0bounds}. 
As mentioned above, in our analysis we use the full Higgs couplings
obtained with  
{\tt FeynHiggs} rather than the approximate formula given in 
\refeq{eq:d0formula}. Similar results would be
obtained with {\tt CPsuperH}.

The impact on the limits in the $\MA$--$\tb$ plane from varying $\mu$ 
while keeping all other parameters fixed can
easily be read off from \refeq{eq:d0formula}. 
For a given value of the $\cp$-odd mass and
$\tb$, the bound on $\sigma(b\bar{b}\phi) \times 
\br (\phi \to b\bar{b})$ provides an upper bound on the 
bottom-quark Yukawa coupling. The main effect therefore is that
as $\mu$ varies, the bound on $\tb$ also changes in
such a way that the value of the bottom Yukawa coupling
at the boundary line in the
$\MA$--$\tb$ plane remains the same. 

\begin{figure}
\BC
\includegraphics[width=.49\textwidth]{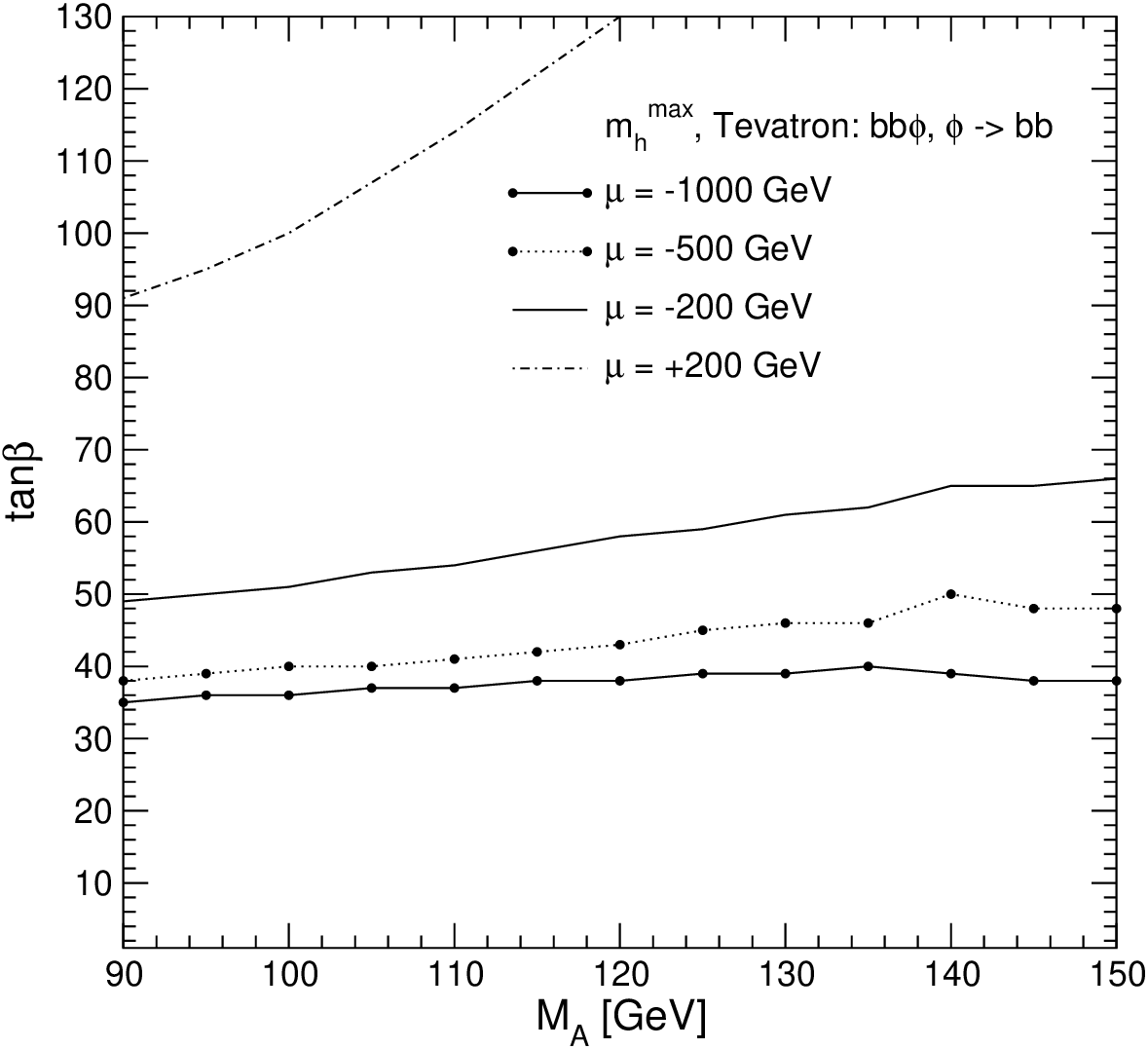}
\includegraphics[width=.49\textwidth]{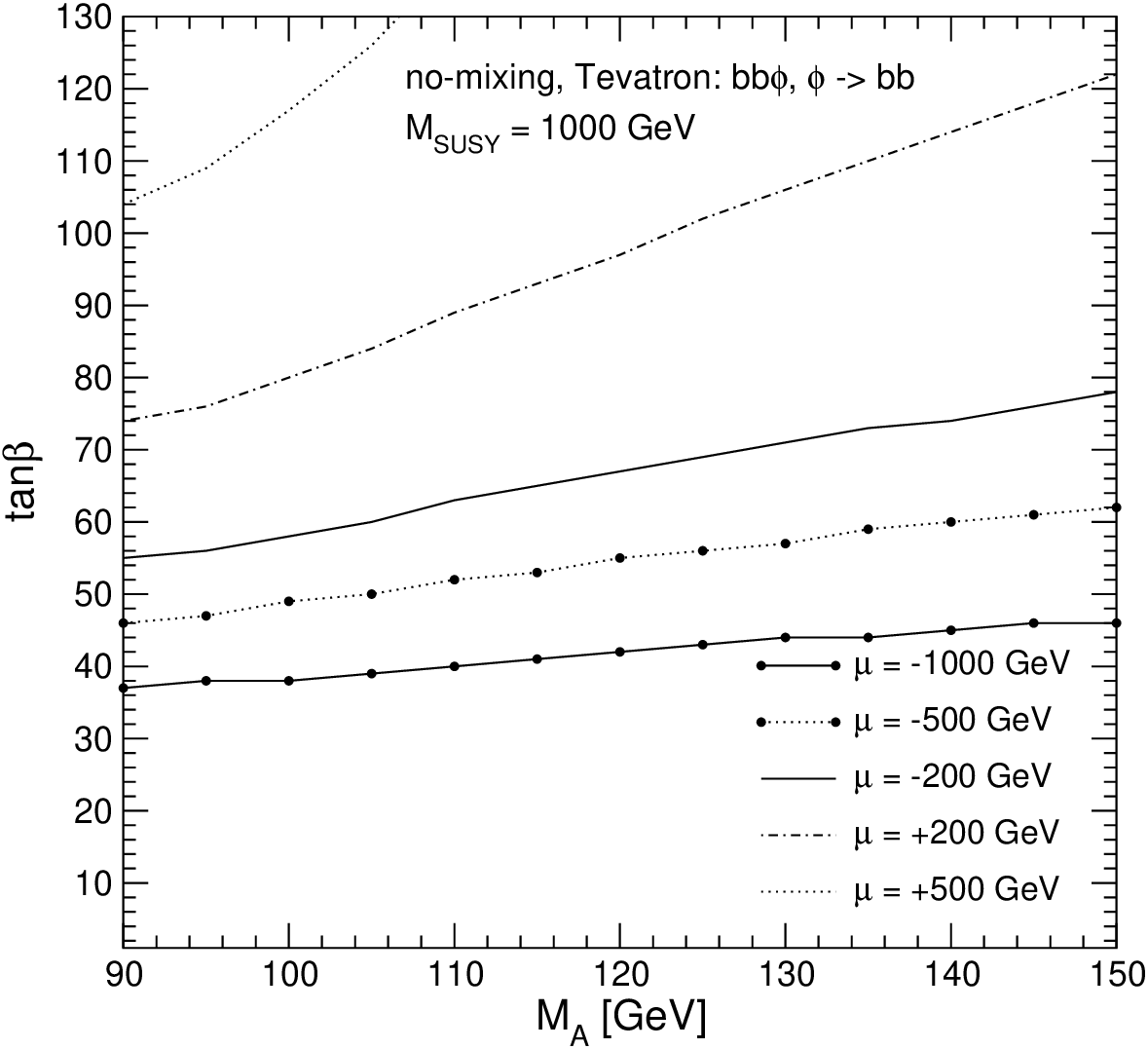}
\EC
\caption{Change in the limits obtained from the $b \bar b \phi$, 
$\phi \to b \bar b$ channel in the $\mhmax$ (left) and no-mixing (right)
benchmark scenarios for different values of $\mu$. The value 
$\mu = -200 \gev$ was chosen by the D0 Collaboration in
\citere{D0bounds}. 
The other curves indicate the corresponding limits
for $\mu = +200, \pm 500, \pm 1000 \gev$. The curves for 
$\mu = +500, +1000\gev$ ($\mu = +1000 \gev$)
do not appear in the left (right) plot for the $\mhmax$ (no-mixing) 
scenario, since for
these $\mu$ values there is no $\tb$ exclusion below $\tb = 130$
for any value of $\MA$.
}
\label{fig:d0change}
\end{figure}

The dependence of the limits in the $\MA$--$\tb$ plane obtained
from the process $b \bar b \phi, \phi \to b \bar b$ on the parameter $\mu$
is shown in \reffi{fig:d0change}. 
The limits for $\mu = -200\gev$ in the 
$\mhmax$ and no-mixing scenarios, corresponding to the limits presented
by the D0 Collaboration in \citere{D0bounds}, are compared with the limits
arising for different $\mu$ values,
$\mu = +200, \pm 500, \pm 1000 \gev$.
\reffi{fig:d0change} illustrates that the effect
of changing the sign of $\mu$ on the limits in the $\MA$--$\tb$ plane
obtained from the process $b \bar b \phi, \phi \to b \bar b$ is quite
dramatic. In the $\mhmax$ scenario the exclusion bound degrades from 
about $\tb = 50$ for $\MA = 90\gev$ in the case of $\mu = -200\gev$ to
about $\tb = 90$ for $\MA = 90\gev$ in the case of $\mu = +200\gev$.
We extend our plots to values of $\tb$ much larger than 50 mainly for
illustration purposes; the region $\tb \gg 50$ in the MSSM is theoretically
disfavoured, if one demands that the values of the bottom and $\tau$
Yukawa couplings remain in the perturbative regime up to energies of
the order of the unification scale. The situation for the
bottom-Yukawa coupling can be ameliorated for large positive values of
$\mu$ due to the $\db$ corrections.
The curves for $\mu = +500, +1000\gev$ 
do not appear in the plot for the $\mhmax$ scenario, since for
these $\mu$ values there is no $\tb$ exclusion below $\tb = 130$
for any value of $\MA$. On the other hand, the
large negative values of $\mu$ shown in \reffi{fig:d0change}, 
$\mu = -500, -1000\gev$,
lead to an even stronger
enhancement of the signal cross section than for $\mu = -200 \gev$ and,
accordingly, to an improved reach in $\tb$. It should be noted that for 
$\mu = -500, -1000\gev$ the bottom Yukawa coupling becomes so large for
$\tb \gg 50$ that a perturbative treatment would no longer be reliable
in this region.

In the no-mixing scenario, where the absolute value of
$\db$ is smaller, the exclusion bound is shifted from 
about $\tb = 55$ for $\MA = 90\gev$ in the case of $\mu = -200\gev$ to
about $\tb = 75$ for $\MA = 90\gev$ in the case of $\mu = +200\gev$.
For $\mu = +500\gev$, no excluded region can be established below
$\tb = 100$.
As above, large negative values of $\mu$, i.e.\
$\mu = -500, -1000\gev$, result in an improved
reach in $\tb$ compared to the value $\mu = -200\gev$
chosen by the D0 Collaboration.

The variation with the sign and absolute value of the $\mu$-parameter shows
the strong dependence of the limits in the $\MA$--$\tb$ plane on
the strength of the bottom-quark Yukawa coupling and hence on the
supersymmetric parameter space. 
The sensitive dependence of the 
process $b \bar b \phi, \phi \to b \bar b$ on the bottom-quark Yukawa
coupling is not specific to the particular benchmark scenarios
considered here. Keeping the ratio of $\mu \mgl/\msusy^2$ fixed but 
varying $\mu$ and $\mgl$ independently will lead to similar results as
those shown here. 
A scenario where large compensations are possible between the two
contributions entering $\db$, see \refeq{def:dmb}, will be discussed
below.
Scenarios with different values of the other supersymmetric
parameters (besides the ones entering $\db$) will reproduce a similar
behaviour as those discussed here.

In \citere{benchmark2} the definition of the $\mhmax$ and no-mixing
scenarios given in \citere{benchmark} has been updated, and the 
``small~$\aeff$''~scenario and the ``gluophobic Higgs scenario'' have
been proposed as additional scenarios for the search for the light
$\cp$-even Higgs boson at the Tevatron and the LHC. The sign of $\mu$ in
the $\mhmax$ and no-mixing scenarios has been reversed to
$\mu = +200\gev$ in 
\citere{benchmark2}. 
This leads typically to a better agreement with the constraints from 
$(g - 2)_{\mu}$.
Furthermore, the value of $\msusy$ 
in the no-mixing scenario was increased from $1000 \gev$~\cite{benchmark}
to $2000 \gev$ in order to ensure that most of the parameter space of this
scenario is in accordance with the LEP exclusion
bounds~\cite{LEPHiggsSM,LEPHiggsMSSM}. 

Another scenario defined in \citere{benchmark2} is the 
``constrained-$\mhmax$'' scenario. It differs from the $\mhmax$~scenario
as specified in \citere{benchmark2} by the reversed sign of $\Xt$,
\BEA
\Xt^{\OS} &=& -2\, \msusy  \; \mbox{(FD calculation)}, \non \\
\Xt^{\MS} &=& -\sqrt{6}\, \msusy \; \mbox{(RG calculation)}, \non \\
\mu &=& +200 \gev ~.
\label{constrainedmhmax}
\EEA
For small $\MA$ and minimal flavor violation this results in better
agreement with the constraints from  
$\br(b \to s \ga)$. For large $\tb$ one has $\At \approx \Xt$, thus
$\At$ and $\mgl$ have opposite signs. This can lead to cancellations in
the two contributions entering $\db$, see \refeq{def:dmb}. 
In contrast to the $\mhmax$ scenario, where the two contributions
entering $\db$ add up, see \refeq{eq:db1}, the constrained-$\mhmax$
scenario typically 
yields relatively small values of $\db$ and therefore a correspondingly
smaller effect
on the relation between the bottom-quark mass
and the bottom Yukawa coupling, e.g.\
\BEA
\mbox{constrained-$\mhmax$ scenario, $\mu = +200 \gev$, $\tb = 50$} &:&
\db = -0.001~. \label{eq:db3} 
\EEA
For large values of $|\mu|$ the compensations between the two terms
entering $\db$ are less efficient, since the function $I$ in the second
term of  \refeq{eq:db1} scales like $1/\mu^2$ for large $|\mu|$.

\begin{figure}[htb!]
\BC
\includegraphics[width=.49\textwidth]{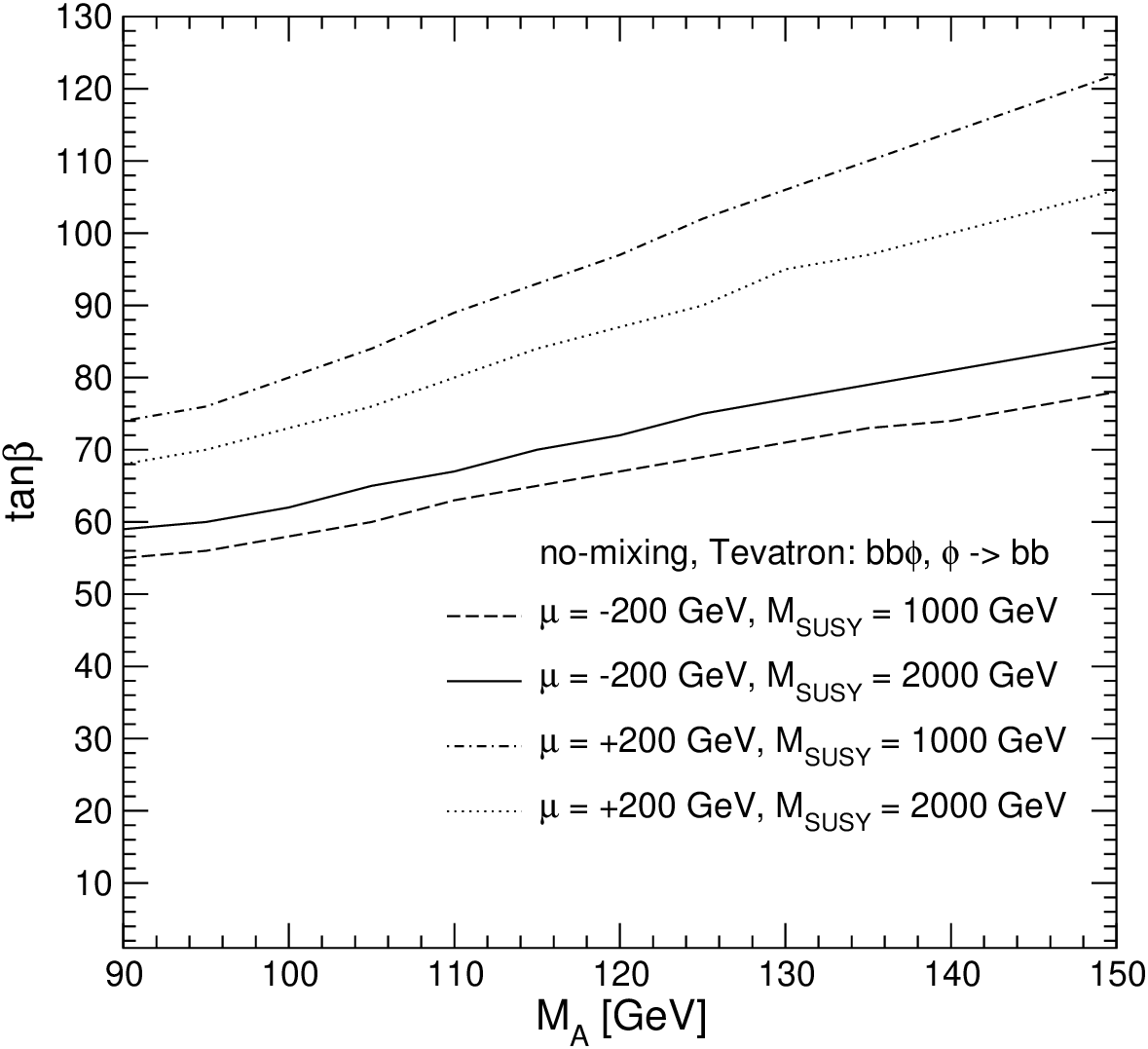}
\includegraphics[width=.49\textwidth]{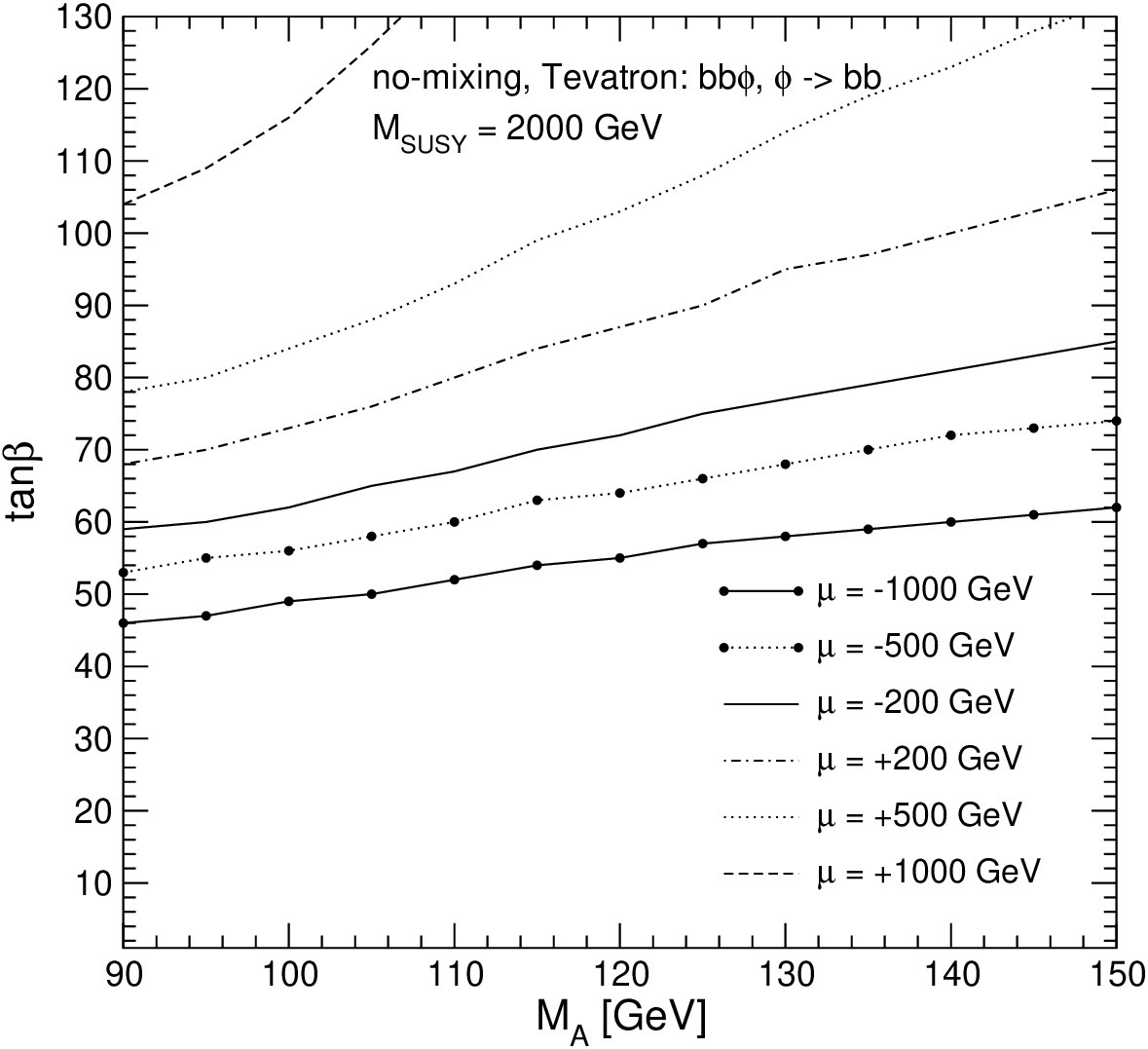}
\EC
\caption{Variation of the limits obtained from the $b \bar b \phi$, 
$\phi \to b \bar b$ channel in the no-mixing scenario for 
different values of $\msusy$ and $\mu$. 
The left plot shows the results for $\msusy = 1000, 2000 \gev$ and 
$\mu = \pm 200 \gev$, while in the right plot the results
for $\msusy = 2000 \gev$ and 
$\mu = \pm 200, \pm 500, \pm 1000 \gev$
are given. 
}
\label{fig:d0change2}
\end{figure}

We now study the impact of the benchmark definitions of
\citere{benchmark2} on the limits in the $\MA$--$\tb$ plane arising from
the $b \bar b \phi$, $\phi \to b \bar b$ channel.
The left plot in \reffi{fig:d0change2} shows the effect of
changing $\msusy = 1000 \gev$ to $\msusy = 2000 \gev$ in the no-mixing
scenario for $\mu = \pm 200 \gev$. Due to the heavier scalar bottoms
in the case of $\msusy = 2000 \gev$ the effect of the $\db$ corrections 
is suppressed as compared to the benchmark definition in
\citere{benchmark}. This leads to a shift of the limits in the
$\MA$--$\tb$ plane of about $\De\tb = 5$--10 for a given value of $\MA$.
The right plot of \reffi{fig:d0change2} shows for $\msusy = 2000 \gev$ 
the variation of the limits with $\mu$.
In this case even for
$\mu = +1000 \gev$ a $\tb$ exclusion limit can be established below 
$\tb = 130$, in contrast to the scenario with $\msusy = 1000 \gev$, 
see~\reffi{fig:d0change}. 

\begin{figure}[htb!]
\BC
\includegraphics[width=.49\textwidth]{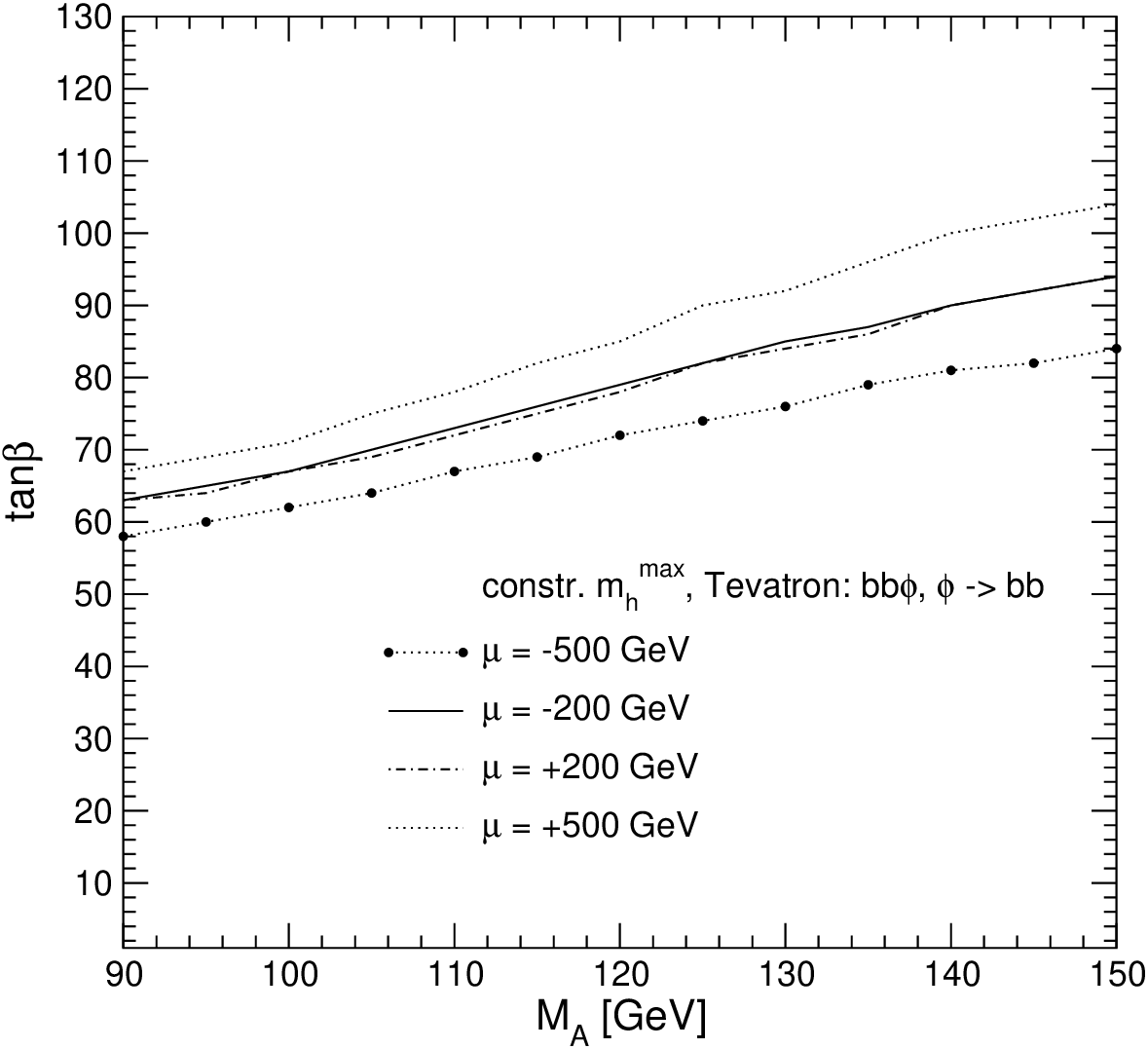}
\includegraphics[width=.49\textwidth]{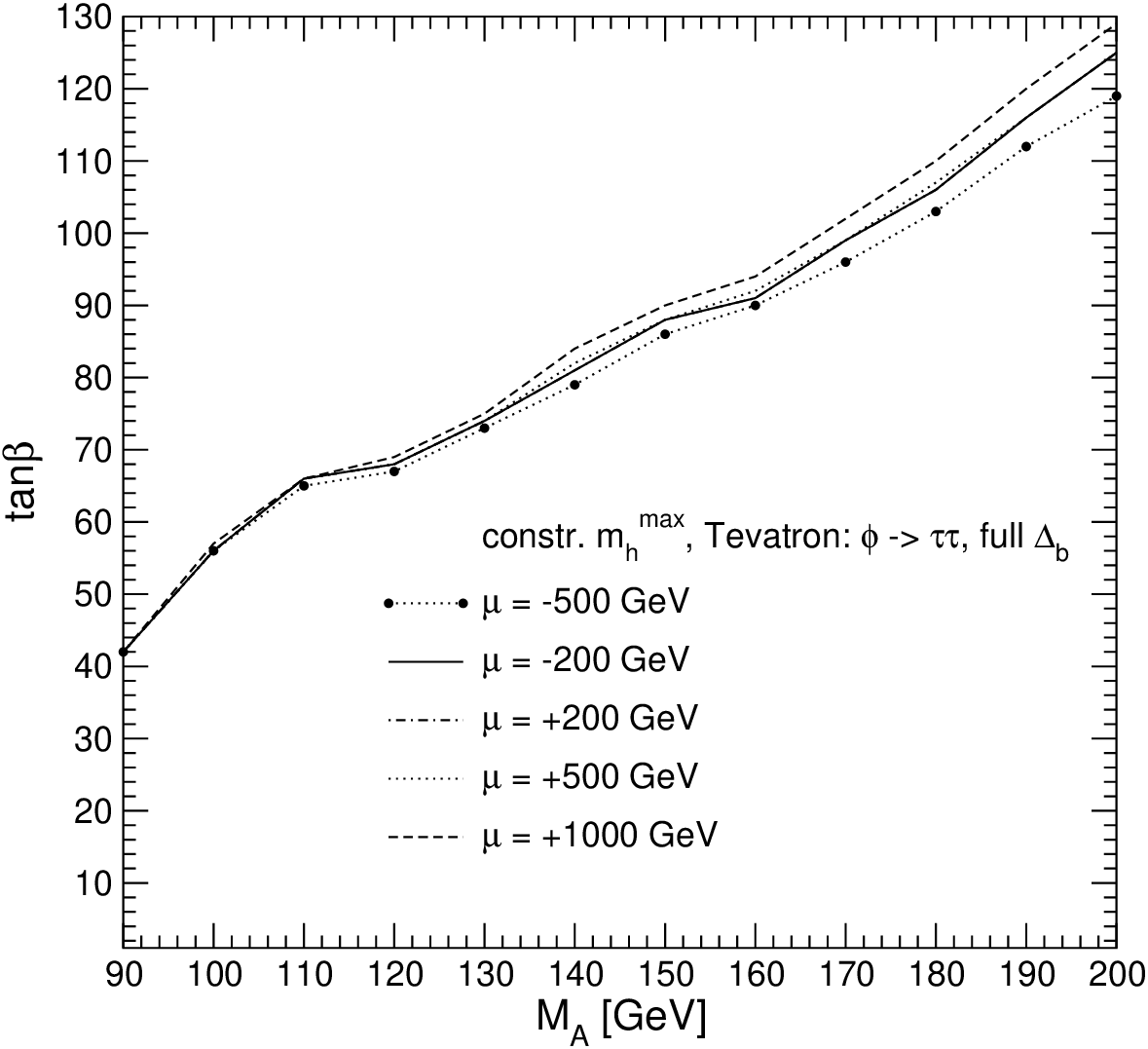}
\EC
\caption{Left: Variation of the limits obtained from the $b \bar b \phi$, 
$\phi \to b \bar b$ channel in the constrained-$\mhmax$ scenario for 
different values of $\mu$.
Right: Variation of the limits obtained from the 
$p \bar p \to \phi \to \tau^+\tau^-$ channel in the constrained-$\mhmax$
scenario for different values of $\mu$.
}
\label{fig:constmhmaxTev}
\end{figure}

The results in the constrained-$\mhmax$ scenario are displayed in
\reffi{fig:constmhmaxTev} (left). The results are shown for 
$\mu = \pm 200, \pm 500 \gev$. 
As expected from the discussion above, the obtained limits are
relatively stable against the variation of $\mu$. 
For $\mu = +500 (-500) \gev$ the $\tb$ limit is significantly weaker
(stronger) than for
smaller values of $|\mu|$ as a consequence of the less efficient
cancellation of the two contributions to $\db$ discussed above.
Nevertheless, the limits obtained for $|\mu| \le 500 \gev$ are
weaker than those for the $\mhmax$ scenario with negative $\mu$, but
stronger than those for positive $\mu$.
The curve for $\mu = +1000\gev$ is not shown in the
plot, since for
this value there is no $\tb$ exclusion below $\tb = 130$
for any value of $\MA$. 
For $\mu = -1000 \gev$, on the other hand,
the radiative corrections lead to a large mass splitting between the
$\cp$-odd and $\cp$-even Higgs boson masses so that the approximation of
adding the two signal cross sections is no longer valid, see the
discussion in \refse{sec:impact}.
A more detailed study would be necessary to incorporate also the case
of larger Higgs boson mass splittings.

\subsubsection{Limits from the process $p \bar p \to \phi \to \tau^+\tau^-$}

\label{sec:CDFtautau}

The limits obtained from the $p \bar p \to \phi \to \tau^+\tau^-$
channel by the CDF Collaboration were presented in the $\MA$--$\tb$
plane for the $\mhmax$ and no-mixing scenarios as defined in
\citere{benchmark2} and employing two values of the $\mu$ parameter,
$\mu = \pm 200 \gev$. According to the discussion in \refse{sec:impact},
the limits obtained from the $p \bar p \to \phi \to \tau^+\tau^-$
channel are expected to show a weaker dependence on the sign and
absolute value of $\mu$ than the limits arising from the 
$b \bar b \phi, \phi \to b \bar b$ channel. 
On the other hand, for large values of $\tb$ and
negative values of $\mu$, 
the large corrections to the bottom Yukawa coupling discussed above can
invalidate a perturbative treatment for this channel.

The MSSM prediction for 
$\si(p \bar p \to \phi) \times {\rm BR} (\phi \to \tau^+\tau^-)$ as 
a function of $\tb$ has been evaluated by the CDF collaboration using
the {\tt HIGLU} program~\cite{higlu} for the gluon fusion channel.
The prediction for the $b \bar b \to \phi + X$ channel was obtained from the
NNLO result in the SM from \citere{robikilgore}, and 
$\left[\si \times {\rm BR}\right]_{\rm MSSM} /
\left[\si \times {\rm BR}\right]_{\rm SM}$ was calculated with the 
{\tt FeynHiggs} 
program~\cite{feynhiggs,feynhiggs1.2,feynhiggs2,feynhiggs2.3}. While
the full $\db$ correction to the bottom Yukawa
correction was taken into 
account in the $b \bar b \to \phi + X$ production channel and the 
$\phi \to \tau^+\tau^-$ branching ratios, the public version of the 
{\tt HIGLU} program~\cite{higlu} does not include the $\db$ correction for
the bottom Yukawa coupling entering the bottom loop contribution to the 
$gg \to \phi$ production process. In order to 
treat the two contributing production processes in a uniform way, 
the $\db$ correction should be included (taking into account the
\order{\als} and the \order{\alt} parts, see \refeq{def:dmb}) in the 
$gg \to \phi$ production process 
calculation. For the large value of $\msusy$ chosen 
in the $\mhmax$ and no-mixing benchmark scenarios other higher-order
contributions involving sbottoms and stops can be 
neglected (these effects are small provided $\msusy \gsim 500$~GeV).

We therefore begin the investigation of the 
$p \bar p \to \phi \to \tau^+\tau^-$ channel by analyzing the impact of
including or omitting the $\db$ correction in the $gg \to \phi$
production process. In order to get a qualitative understanding of
the variation of the limits on $\tb$ for a given value of $\MA$
induced by the inclusion of the $\db$ corrections in the
gluon fusion channel, it is again useful to employ the simple approximate
formulae given in \refse{sec:impact}. As discussed above, the production 
cross sections may be approximately obtained from the SM ones by including
a simple rescaling by a factor $\tan^2\be/(1 + \db)^2$. 
Hence, defining $\sigma_b$ 
and $\sigma_g$ as the SM cross sections for the $b$-quark associated 
and gluon fusion production of Higgs bosons, respectively, we get
\begin{eqnarray}
\left. \sigma(p \bar{p} \to \phi) \times {\rm BR}(\phi \to \tau^+ \tau^-) 
\right|_{{\rm full} \; \db} 
& \simeq &
\left(\sigma_b + \sigma_g \right) 
\times \frac{\tan^2\be}{(1 + \db)^2 + 9}~, \\
\left.
\sigma(p \bar{p} \to \phi) \times {\rm BR}(\phi \to \tau^+ \tau^-) 
\right|_{{\rm partial} \; \db} 
& \simeq &
\left(\sigma_b + \sigma_g \; (1 + \db)^2 \right) 
\times \frac{\tan^2\be}{(1 + \db)^2 + 9}~, 
\end{eqnarray}
where ``full $\db$'' denotes the case where the $\db$ correction is
incorporated in both the $b$-quark associated and the
gluon fusion production processes (and the $\phi \to \tau^+ \tau^-$
branching ratio), while ``partial $\db$'' denotes the case where the 
$\db$ correction is omitted in the $gg \to \phi$ production process.
The expression above shows that, for positive values of $\mu$, for
which $\db > 0$, the omission of the correction to the $gg \to \phi$
process leads
to an enhancement of the total production cross section with respect to 
the value obtained when these corrections are included. For negative 
values of $\mu$, instead, the situation is reversed.  

The cross section for the $gg \to \phi$ production process including the 
$\db$ correction can be obtained by a simple rescaling of the 
{\tt HIGLU} result for the SM. The cross section for the SM production
rate involving the $b$-quark loop alone is rescaled with
$\Ga(\hbb)_{\rm MSSM}/\Ga(\hbb)_{\rm SM}$, where $\db$ enters the
calculation of $\Ga(\hbb)_{\rm MSSM}$ (SM-QCD corrections to the
$b$-quark mass factorize and drop out in this ratio). As stressed above, 
it has been checked that
the $t$-quark loop gives only a negligible contribution in the MSSM for
$\tb \gsim 20$. The loops involving scalar tops and bottoms, beyond
those included in the $\db$ corrections, also give
small contributions due the heavy scalar masses in the benchmark
scenarios. 

\begin{figure}[htb!]
\BC
\includegraphics[width=.49\textwidth]{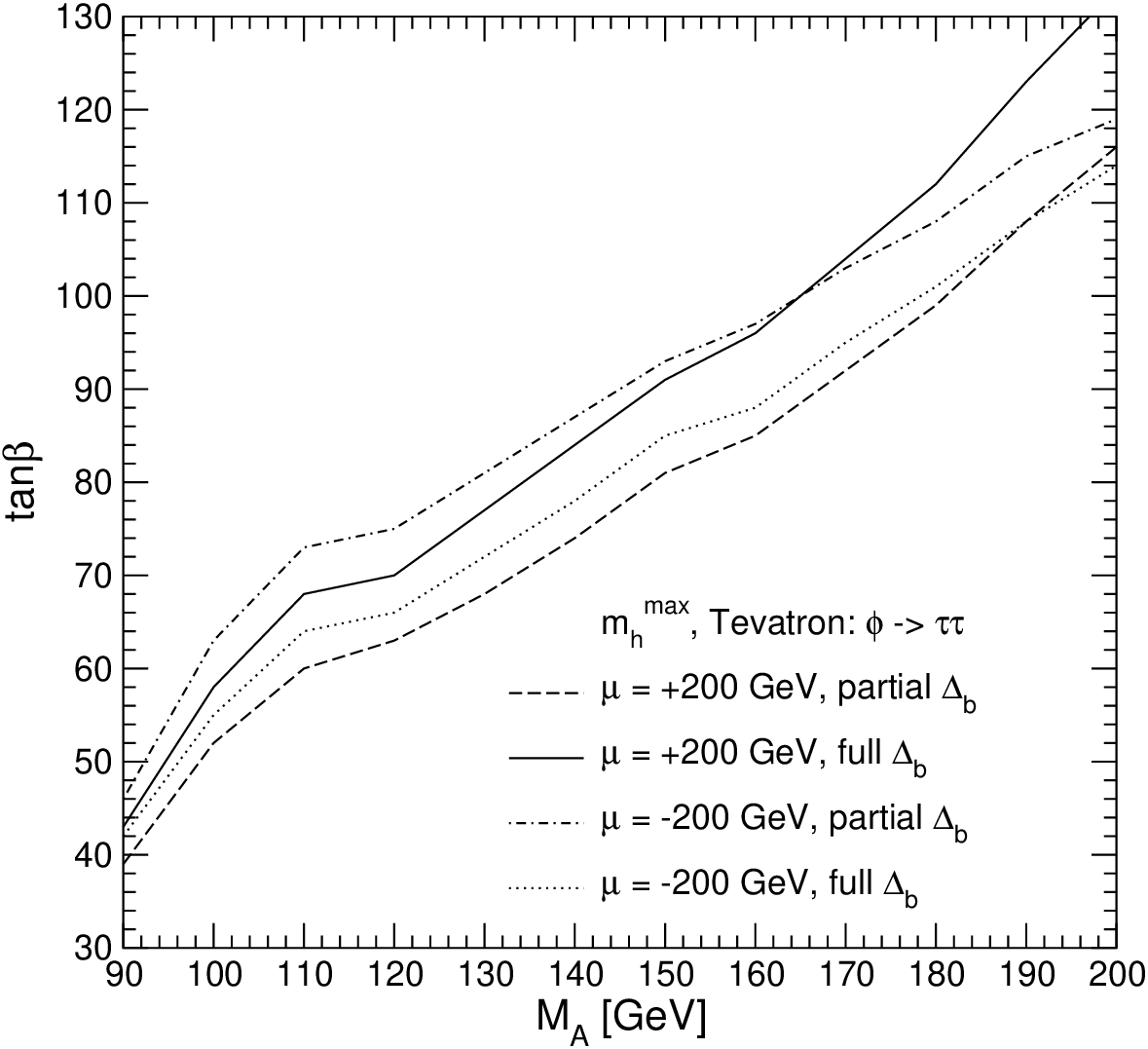}
\includegraphics[width=.49\textwidth]{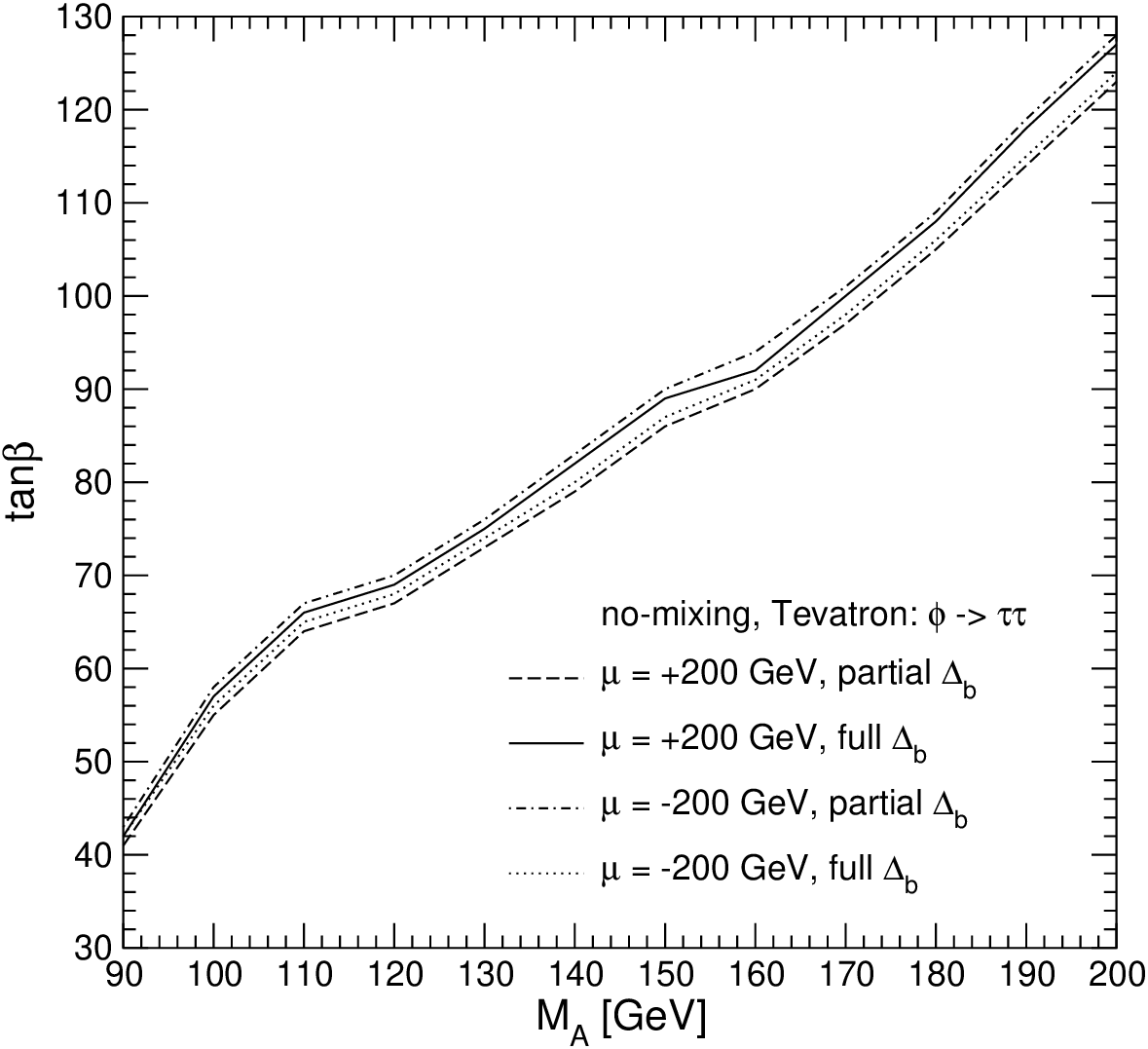}
\EC
\caption{Impact of including or omitting the $\db$ correction in the 
$g g \to \phi$ production process on the limits obtained from the 
$p \bar p \to \phi$, $\phi \to \tau^+\tau^-$ channel. The results are shown
for $\mu = \pm 200 \gev$ in the $\mhmax$ (left) and no-mixing
(right) benchmark scenarios~\cite{benchmark2}.}
\label{fig:cdfdeltab}
\end{figure}

The comparison of the ``partial $\db$'' and the ``full $\db$'' results
is shown in \reffi{fig:cdfdeltab}. The evaluation of $\si \times \br$
in the MSSM has been performed by using {\tt FeynHiggs} for
rescaling the {\tt HIGLU} results for the $gg \to \phi$
production cross sections and the SM results for the $b \bar b \to \phi + X$
channel~\cite{robikilgore} by the appropriate MSSM correction factor, as
outlined above. The impact of the additional contribution can be read off
from \reffi{fig:cdfdeltab} by comparing the 
results where the $\db$ corrections are
omitted in the $gg \to \phi$ production cross sections (``partial
$\db$'') with the results where the $\db$ corrections have been taken
into account everywhere in the production and decay processes (``full
$\db$''). The effect on the exclusion bounds in the $\MA$--$\tb$ plane
is seen to be quite significant. While in the case where the $\db$
corrections are neglected in the $gg \to \phi$ production cross sections
the strongest exclusion bounds are obtained for positive values of
$\mu$, the inclusion of the $\db$ corrections reverses this situation.
As explained above, the inclusion of the $\db$ corrections to the 
$gg \to \phi$ production process leads to a larger cross section (and
correspondingly to a stronger $\tb$ bound) in the case of negative
$\mu$, while the cross section is suppressed for positive values of $\mu$.
The corresponding shifts of the exclusion limits in the $\MA$--$\tb$
plane amount up to $\De\tb \sim 10$ for the $\mhmax$ scenario. In the 
no-mixing sceanrio (defined according to \citere{benchmark2}) the effect
is less pronounced because of the smaller numerical value of $\db$, 
giving rise to shifts in the exclusion limits up to $\De\tb \sim  5$.

Following our analysis, the CDF Collaboration has adopted the
prescription outlined above for incorporating the $\db$ correction into
the $gg \to \phi$ production process. The limits given in
\citere{CDFbounds} are based on the MSSM prediction where the $\db$
correction is included everywhere in the production and decay processes
(see e.g.\ \citere{cdfold} for a previous analysis).

\begin{figure}[htb!]
\BC
\includegraphics[width=.49\textwidth]{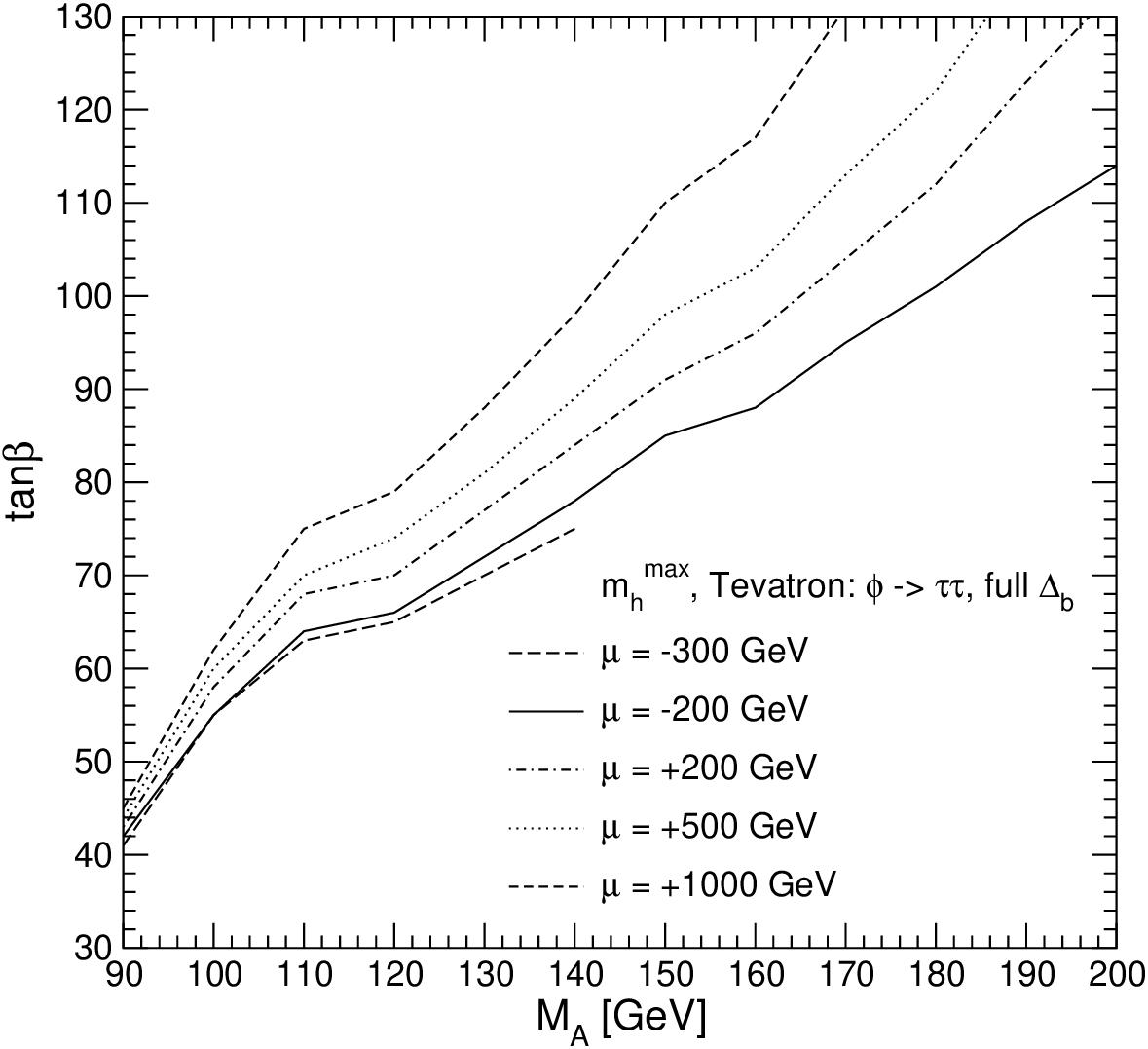}
\includegraphics[width=.49\textwidth]{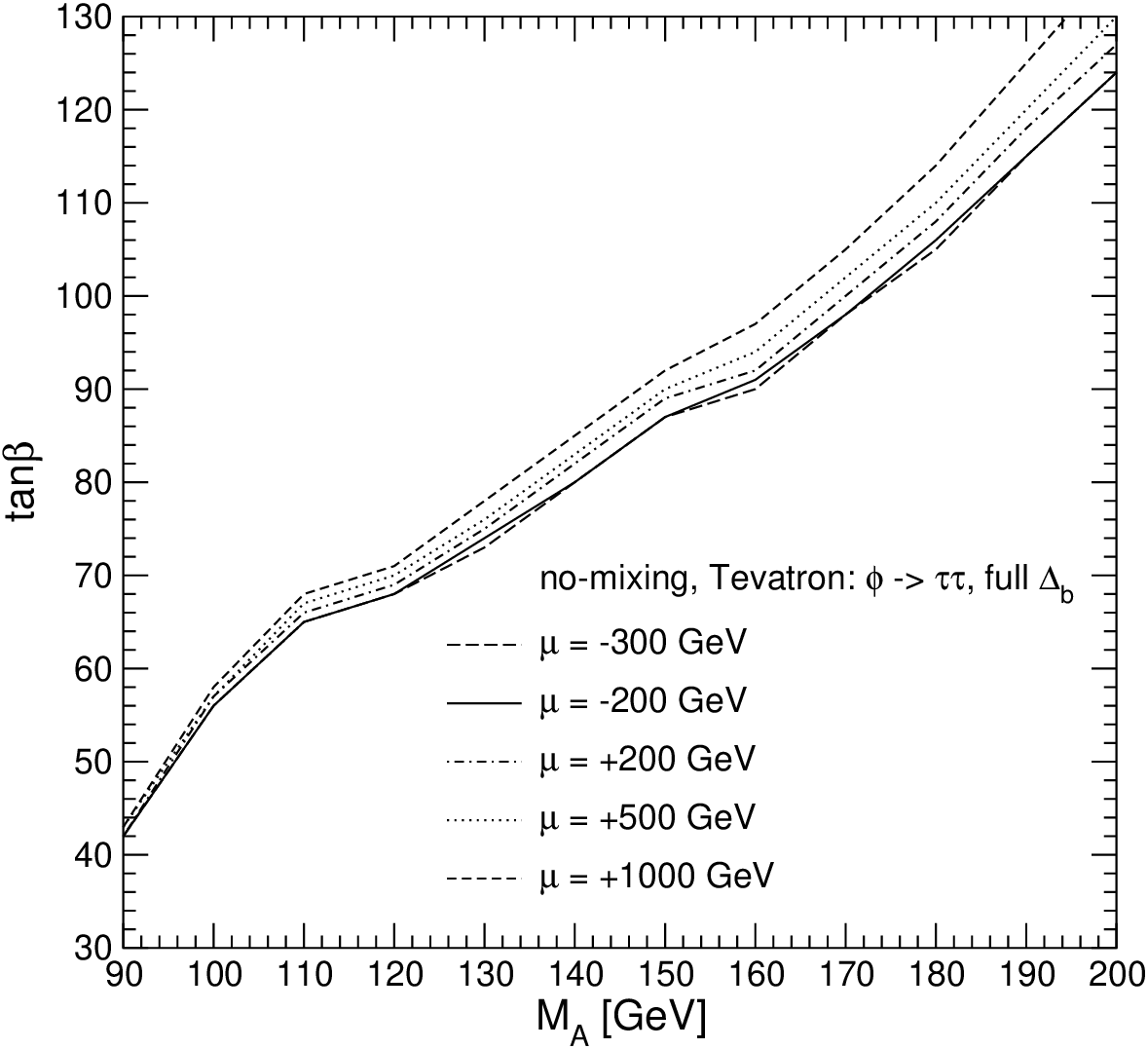}
\EC
\caption{Variation of the limits obtained from the 
$p \bar p \to \phi \to \tau^+\tau^-$ channel at the Tevatron 
in the $\mhmax$ (left) and 
no-mixing (right) benchmark scenarios for different values of $\mu$.}
\label{fig:cdfchange}
\end{figure}

We next turn to the discussion of the sensitivity of the limits obtained 
from the $p \bar p \to \phi \to \tau^+\tau^-$ channel (including the $\db$
correction in all production and decay processes) on the sign and
absolute value of $\mu$. As discussed above, similar variations in the 
exclusion limits will occur if the absolute values of $\mu$, $\mgl$, $\At$ 
and $\msusy$ are varied, while keeping the ratios appearing in 
$\db$ constant.  
The results are given in
\reffi{fig:cdfchange} for the $\mhmax$ scenario (left) and the
no-mixing scenario (right). In the $\mhmax$ scenario we find a
sizable dependence of the $\tb$ bounds on the sign and absolute value of
$\mu$.%
\footnote{
For $\mu = -300 \gev$ the curve stops at around $\tb = 75$ because 
the bottom Yukawa coupling becomes very large, leading to 
instabilities in the calculation of the Higgs properties.
For the same reason, even more negative values of $\mu$ are not
considered here.
}
The effect grows with $\MA$ and, 
for the
range of parameters explored in \reffi{fig:cdfchange},
leads to a variation
of the $\tb$ bound larger than  $\De\tb \sim 30$. 
In the
no-mixing scenario the effect is again smaller, but it can still 
lead to a variation of the $\tb$ bounds by as much as
$\De\tb \sim 10$.  

The results obtained in the constrained-$\mhmax$ scenario are shown in
\reffi{fig:constmhmaxTev} (right). As expected, the exclusion limits in
this scenario are very robust with respect to varying $\mu$.
All values of $\mu$ result
practically in the same $\tb$ exclusion bounds. The lines not visible
in the plot are actually covered by a line of another $\mu$~value.
For $\mu = -1000 \gev$, 
the radiative corrections lead to a large mass splitting between the
$\cp$-odd and $\cp$-even Higgs boson masses, see the discussion above.


\subsubsection{Limits from the process 
$p \bar p \to t \bar t \to H^\pm W^\mp \, b \bar b$,
$H^{\pm} \to \tau \nu_{\tau}$}

For the charged Higgs search channel at the Tevatron~\cite{Tevcharged}, 
$p \bar p \to t \bar t \to H^\pm W^\mp b \bar b$,
$H^{\pm} \to \tau \nu_{\tau}$
the variation of the cross section with $\mu$ is driven by the impact
of the $\db$ correction on ${\rm BR}(t \to H^{\pm} b)$~\cite{deltamb2}. 
The decay width $\Gamma(t \to H^{\pm} b)$ is proportional to 
$\tan^2\be/(1 + \db)^2$, leading to an expression for the branching ratio
in analogy to \refeq{eq:BRAbb}. Accordingly, a positive $\db$ leads to a
suppression of ${\rm BR}(t \to H^{\pm} b)$, while a negative $\db$ leads
to an enhancement.

For a fixed value of $\MHp$, the value of $\MA$ is driven to rather
small values because of the tree-level relation
$\MA^2 = \MHp^2 - \MW^2$.
For large values of $\tb$, $\At$ and $\Ab$ this effect is further
enhanced by the higher-order
corrections in \refeq{eq:MHpmHO}. Consequently, 
in the region of small $\MHp$ and large $\tb$ currently
probed at the Tevatron~\cite{Tevcharged} the corresponding $\MA$ values 
tend to be further reduced with respect to the already small
tree-level values%
\footnote{
This effect is avoided if the parameters
$\At$, $\Ab$ and $\mu$ are such that $A_{tb}$ in
\refeq{eq:Atb} becomes sufficiently large and negative. 
This can be realized, for instance, for values of $\MHp > 120 \gev$ and 
$\At = -\Ab$ in the constrained-$\mhmax$ scenario, for small values
of $\mu$.
}%
~and are in general below the LEP exclusion bound~\cite{LEPHiggsMSSM}. 
Therefore, this
channel at present is less relevant for obtaining exclusion limits in
the $\MA$--$\tb$ plane than the neutral Higgs-boson search channels
discussed above. It is expected to become more competitive, however,
with increasing luminosity collected in Run~II of the Tevatron.


\subsection{Prospects for Higgs sensitivities at the LHC}

The most sensitive channels for detecting heavy MSSM Higgs bosons at the
LHC are the channel $pp \to H/A +X, \, H/A \to \tau^+\tau^-$ (making use of
different decay modes of the two $\tau$ leptons) and the channel 
$t H^{\pm}, H^{\pm} \to \tau \nu_{\tau}$
(for $\MHp \geq \mt$)~\cite{atlastdr,cmshiggs}.
We consider here the parameter region $\MA \gg \MZ$, for which the heavy
states $H$, $A$ are widely separated in mass from the 
light $\cp$-even Higgs boson $h$.
Here and in the following we do not discuss search channels where the
heavy Higgs bosons decay into
supersymmetric particles, which depend very sensitively on the model
parameters~\cite{filip,filip2,cmshiggs}, but we will comment below on how these
decays can affect the searches with bottom-quarks and $\tau$-leptons
in the final state.


\subsubsection{Discovery region for the process 
$pp \to H/A +X, \, H/A \to \tau^+\tau^-$}

To be specific, we concentrate in this section on the analysis carried
out by the CMS Collaboration~\cite{KinNik,cmshiggs}. Similar results for 
this channel have also
been obtained by the ATLAS Collaboration~\cite{atlastdr,atlasnote}.
In order to rescale the SM cross sections and branching ratios, 
the CMS Collaboration has used for the branching ratios the {\tt HDECAY} 
program~\cite{hdecay} and for the production cross sections 
the {\tt HIGLU} program~\cite{higlu} ($gg \to H/A$)
and the {\tt HQQ} program~\cite{hqq} ($gg \to b \bar bH$). In the
{\tt HDECAY} program the $\db$ corrections are partially included 
for the decays of the neutral Higgs bosons (only the \order{\als}
contribution to $\db$ is included, see \refeq{def:dmb}).
The {\tt HIGLU} program (see also the discussion in
\refse{sec:CDFtautau}) and {\tt HQQ}, on the other hand,
do not take into account the
corrections to the bottom Yukawa coupling.%
\footnote{
Since {\tt HQQ} is a leading-order program, non-negligible changes can
also be expected from SM-QCD type higher-order corrections.
}%
~The prospective $5 \si$
discovery contours for CMS 
(corresponding to the upper bound of the LHC ``wedge''
region, where only the light $\cp$-even Higgs boson may be observed
at the LHC) have been presented 
in \citeres{KinNik,cmshiggs}
in the 
$\MA$--$\tb$ plane, for an integrated luminosity of 
30~fb$^{-1}$ and 60~fb$^{-1}$. 
The results
were presented in the $\mhmax$ scenario and for different $\mu$ values, 
$\mu = -200, +300, +500 \gev$. It should be noted that decays of
heavy Higgs bosons into charginos and neutralinos open up for small
enough values of the soft supersymmetry-breaking parameters $M_2$ and $\mu$.
Indeed, the results presented in \citeres{KinNik,cmshiggs} show a 
degradation of the discovery reach in the $\MA$--$\tb$ plane for
smaller absolute values of $\mu$, which is due to
an enhanced branching ratio of $H$,
$A$ into supersymmetric particles, and accordingly a reduced branching
ratio into $\tau$ pairs.

We shall  now study the impact of including the $\db$ corrections into the
production cross sections and branching ratios for different values of
$\mu$. The inclusion of the $\db$ corrections leads to a modification of 
the dependence of the production cross section on $\tb$, as well
as of the branching ratios of the Higgs boson decays into $\tau^+ \tau^-$.
For a fixed value of $\MA$, the results obtained by the CMS
Collaboration for the discovery region in $\tb$
can be interpreted in terms of a cross section limit 
using the approximation of rescaling the SM rate for the 
$pp \to H +X, \, H \to \tau^+ \tau^-$ process by the factor
\BE
\TQb_{\rm CMS} \times \frac{\br(\Htautau)_{\rm CMS} +
\br(\Atautau)_{\rm CMS}}
{\br(\Htautau)_{\rm SM}}~. 
\label{HAtautauold}
\EE
In the above, $\tb_{\rm CMS}$ refers to the value of $\tb$ on the
discovery contour (for a given
value of $\MA$) that was obtained in the analysis of the CMS
Collaboration with
30~fb$^{-1}$~\cite{cmshiggs}.
These $\tb$ values as a function of $\MA$ correspond to
the edge of the area in the $\MA$--$\tb$ plane
in which the signal $pp \to H/A +X, \, H/A \to \tau^+\tau^-$ is visible (i.e.\
the upper bound of the LHC wedge region). The branching ratios
$\br(\Htautau)_{\rm CMS}$ and $\br(\Atautau)_{\rm CMS}$ 
in the CMS analysis have been evaluated with {\tt HDECAY}, 
incorporating
therefore only the gluino-sbottom contribution to $\Delta_b$. 

After including all $\db$ corrections, we evaluate the
$pp \to H / A +X, \, H / A \to \tau^+ \tau^-$ process by rescaling the SM
rate with the new factor, 
\BE
\frac{\TQb_{\rn}}{(1 + \db)^2} \times 
\frac{\br(\Htautau)_{\rn} + \br(\Atautau)_{\rn}}{\br(\Htautau)_{\rm SM}}~,
\label{HAtautaunew}
\EE
where $\db$ depends on $\tb_{\rn}$. The quantities have been
evaluted with {\tt FeynHiggs}, allowing also decays into
supersymmetric particles.  
The resulting shift in the discovery reach for the
$pp \to H/A +X, \, H/A \to \tau^+ \tau^-$ channel can be obtained 
by demanding that \refeq{HAtautauold} and \refeq{HAtautaunew} should
give the same numerical result for a given value of $\MA$.

\begin{figure}[htb!]
\BC
\includegraphics[width=.49\textwidth]{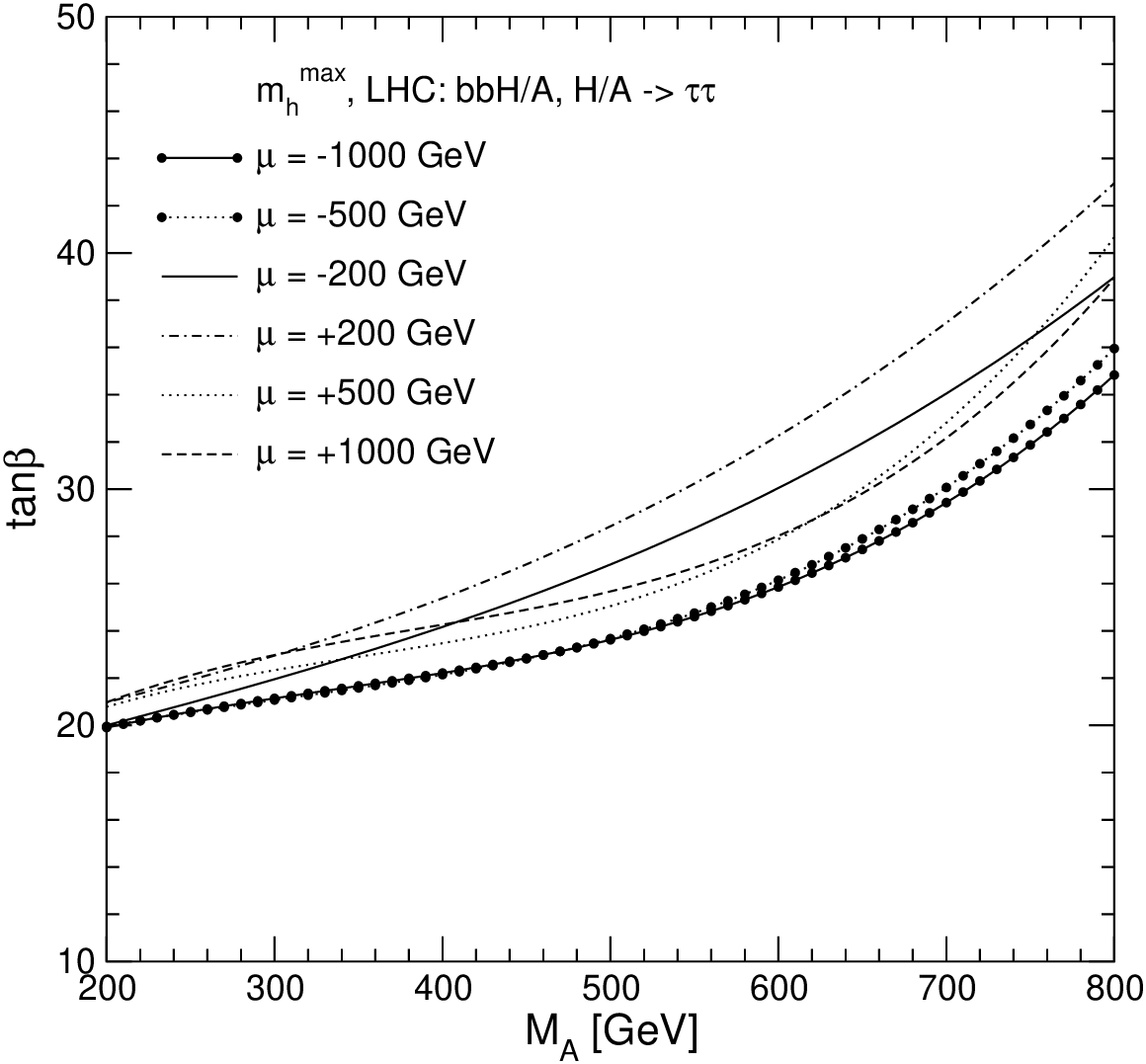}
\includegraphics[width=.49\textwidth]{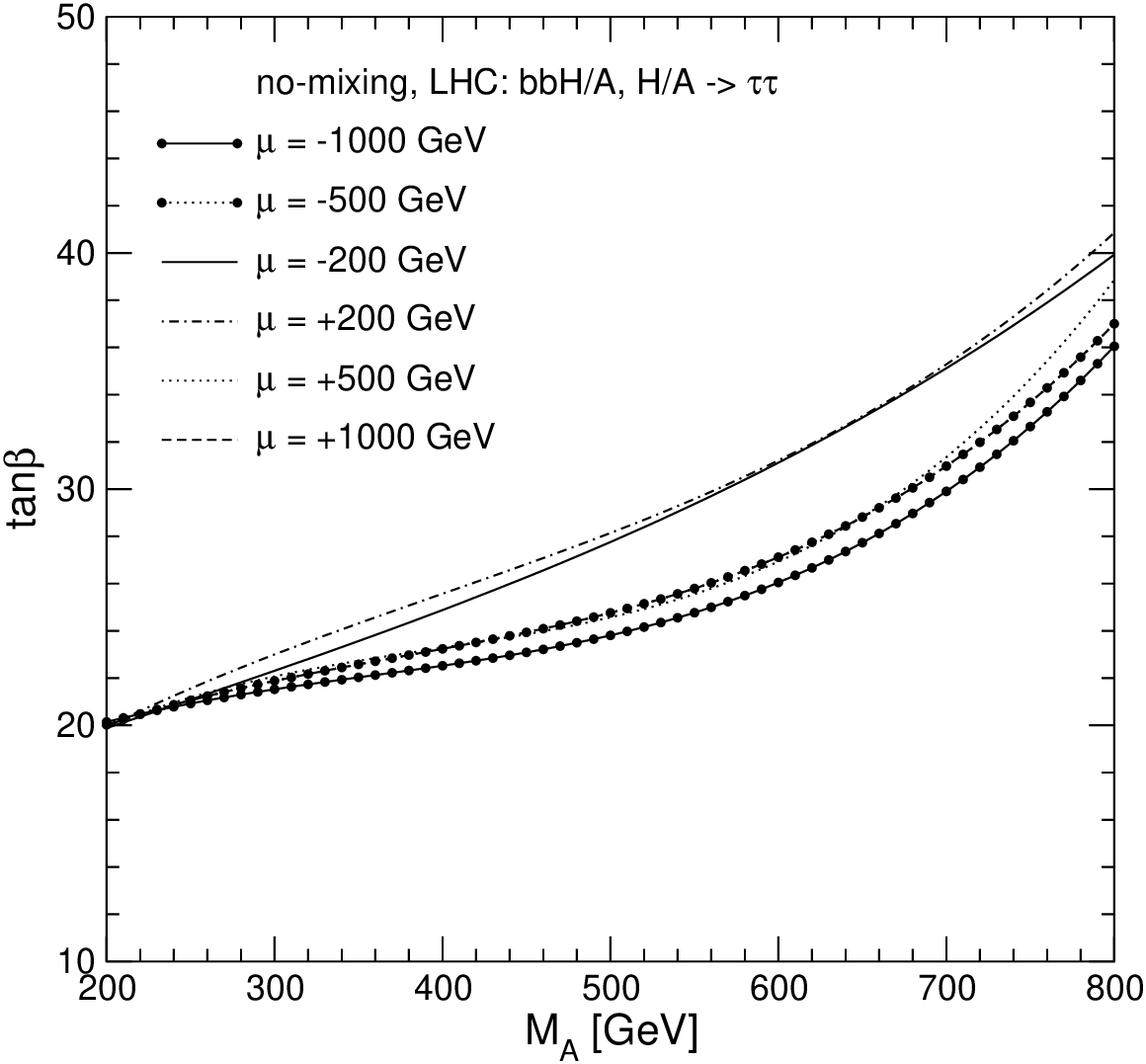}
\EC
\caption{Variation of the $5 \si$ discovery potential for the 
$pp \to H/A +X, \, H/A \to \tau^+\tau^-$ process at the LHC
in the $\mhmax$ (left) and 
no-mixing (right) benchmark scenarios for different values of $\mu$.}
\label{fig:LHCtautau}
\end{figure}

This procedure has been carried out in two benchmark
scenarios for various values of $\mu$. The results are shown in
\reffi{fig:LHCtautau} for the $\mhmax$~scenario (left) and for the
no-mixing scenario (right). The comparison of these results 
with the ones obtained by the CMS Collaboration~\cite{KinNik,cmshiggs} 
shows that
for positive values of $\mu$ the inclusion of the supersymmetric 
radiative corrections leads to a slight shift of the discovery region
towards higher values of $\tb$, i.e.\ to a small increase of the LHC wedge
region. For $\mu = -200 \gev$ the result remains approximately the
same as the one obtained by the CMS Collaboration. 
Due to the smaller considered $\tb$ values compared to the analysis of
the Tevatron 
limits in \refse{sec:CDFtautau}, the corrections to the
bottom Yukawa coupling from $\db$ are smaller, leading to a better
perturbative behavior. As a consequence, also the curves for 
$\mu = -500, -1000 \gev$ are shown in
\reffi{fig:LHCtautau}.

The change in the
upper limit of the LHC wedge region due to the variation of $\mu$ does
not exceed $\De\tb \sim 8$. As explained above, this is a consequence
of cancellations of the leading $\db$ effects
in the Higgs production and the Higgs decay.
Besides the residual $\db$ corrections, a further variation of 
the bounds is caused by the
decays of the heavy Higgs bosons into supersymmetric particles. 
For a given value of $\mu$, the rates of these
decay modes are strongly dependent on the particular values of the
weak gaugino mass parameters $M_2$ and $M_1$. In our analysis,
we have taken $M_2 = 200$~GeV, as established by the benchmark scenarios
defined in \citere{benchmark2}, while $M_1 \simeq 100$~GeV.
Since the Higgs couplings to neutralinos
and charginos depend strongly on the admixture between higgsino and 
gaugino states, the rate of these processes is strongly suppressed for
large values of $|\mu| \gsim 500$~GeV. In general,
the effects of the decays $H/A \to \tilde\chi^0_i \tilde\chi^0_j, 
                                   \tilde\chi^\pm_k \tilde\chi^\mp_l$
only play a role for $\MA \gsim |\mu| + M_1$. 
Outside this range the cancellations of the
$\db$ effects result in a very weak dependence of the rates on $\mu$.

\begin{figure}[htb!]
\vspace{2em}
\BC
\includegraphics[width=.49\textwidth]{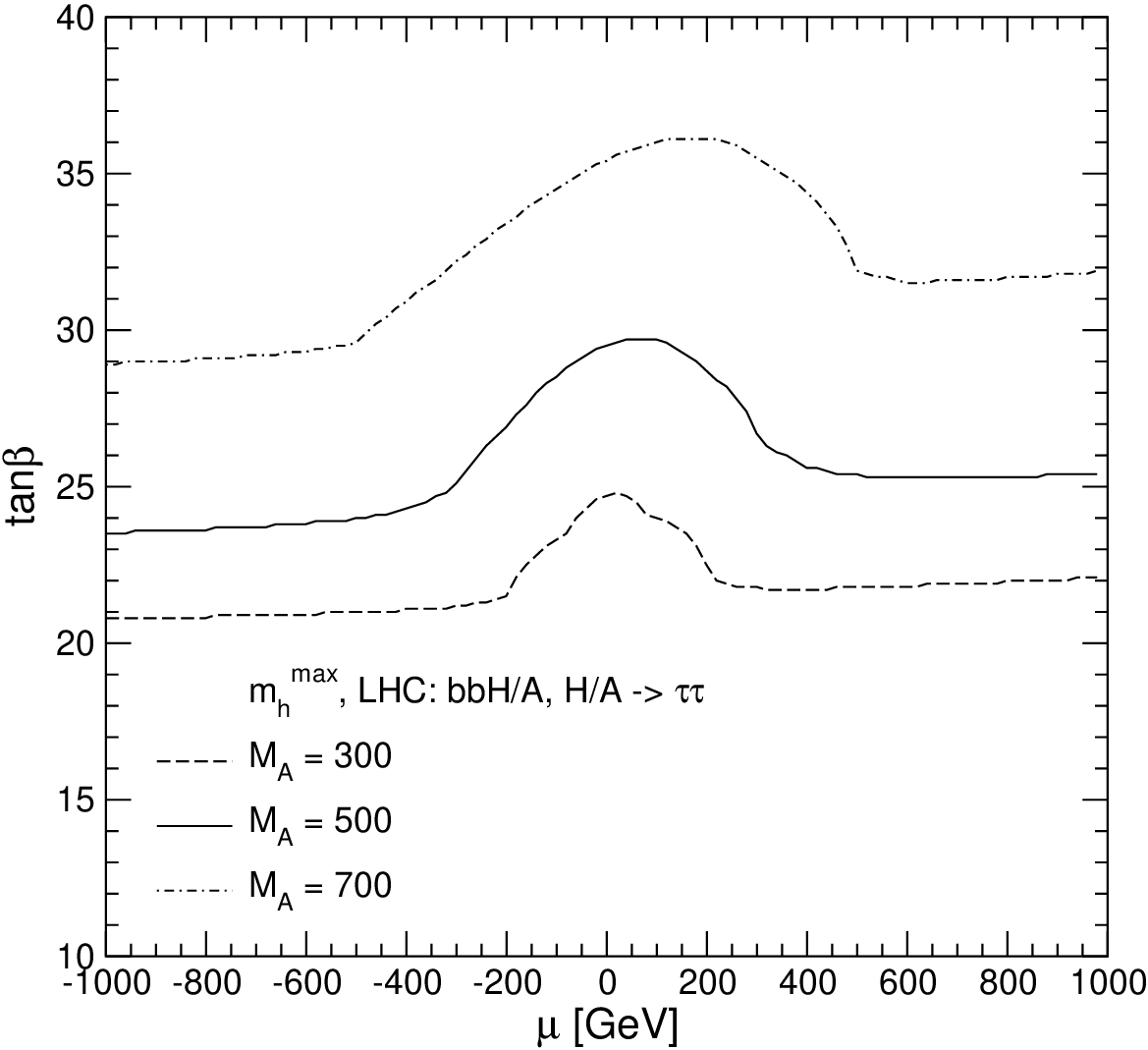}
\includegraphics[width=.49\textwidth]{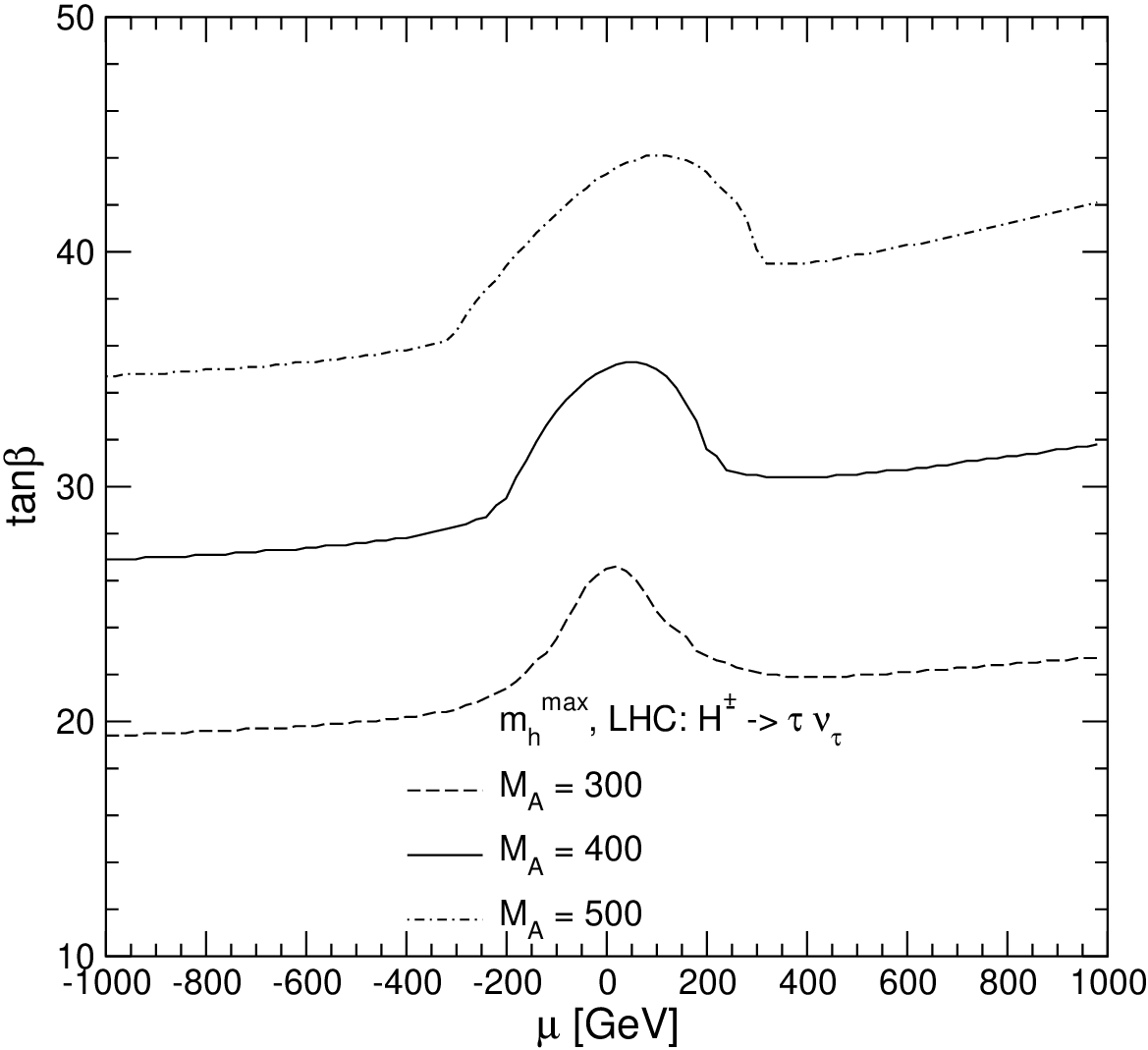}
\EC
\caption{Variation of the $5 \si$ discovery contours 
as a function of the parameter $\mu$ in the $\mhmax$ scenario
for the $pp \to H/A +X, \, H/A \to \tau^+\tau^-$ process (left) and 
the $H^{\pm} \to \tau \nu_\tau$ process (right).}
\label{fig:AchHsusydecay}
\end{figure}

The combination of the effects from supersymmetric radiative corrections
and decay modes into supersymmetric particles gives rise to a rather
complicated dependence of the discovery contour on $\mu$. This is
illustrated in \reffi{fig:AchHsusydecay} (left), where the discovery
contour for the $pp \to H/A +X, \, H/A \to \tau^+\tau^-$ process
is shown as a function of $\mu$ in the $\mhmax$ scenario 
for different values of $\MA$. As explained above, for 
$\MA \gsim |\mu| + M_1$ the decay modes into supersymmetric particles
have a significant impact, while outside this region the dependence on
$\mu$ is rather weak.

\begin{figure}[htb!]
\vspace{3em}
\BC
\includegraphics[width=.49\textwidth]{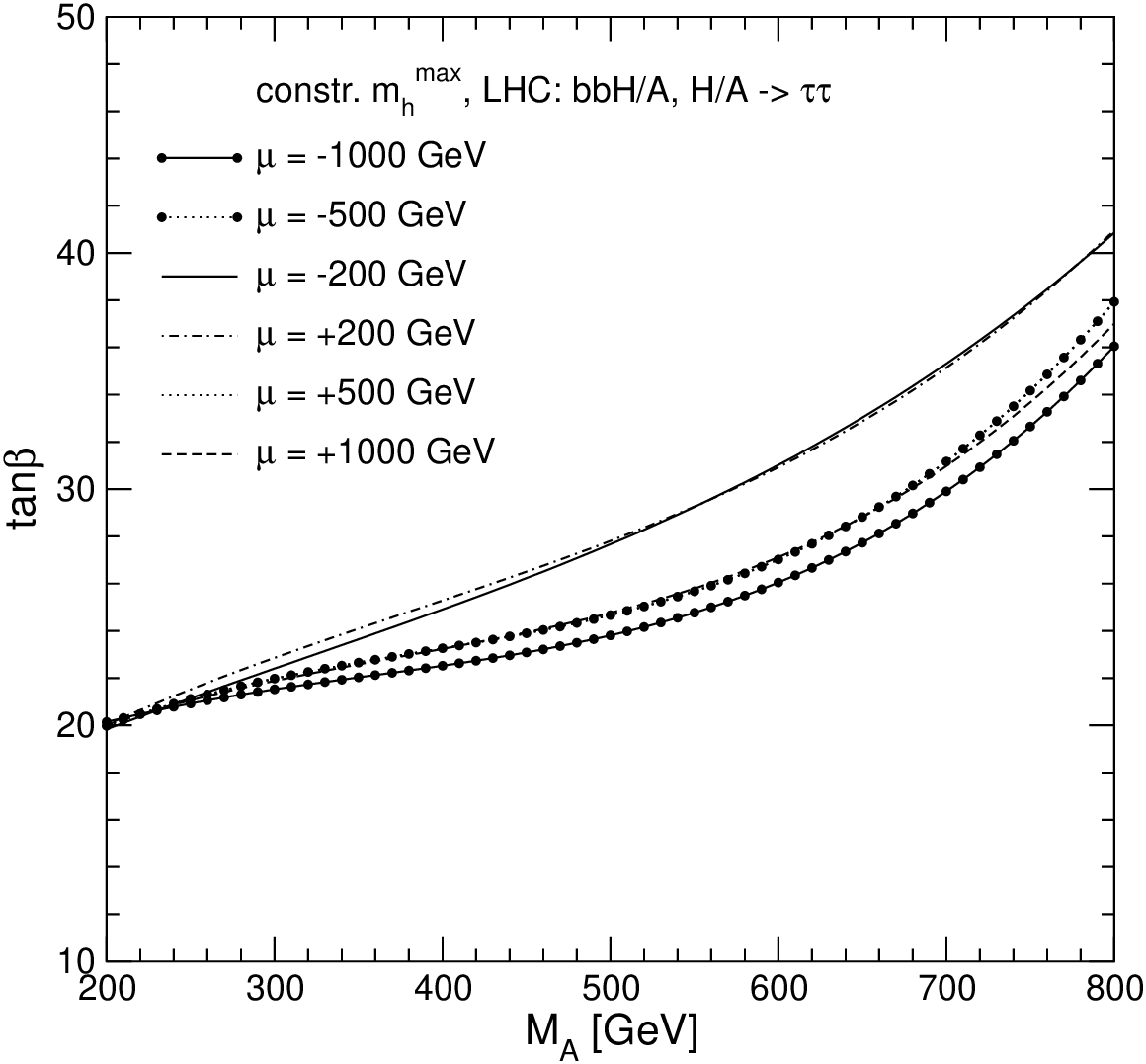}
\includegraphics[width=.49\textwidth]{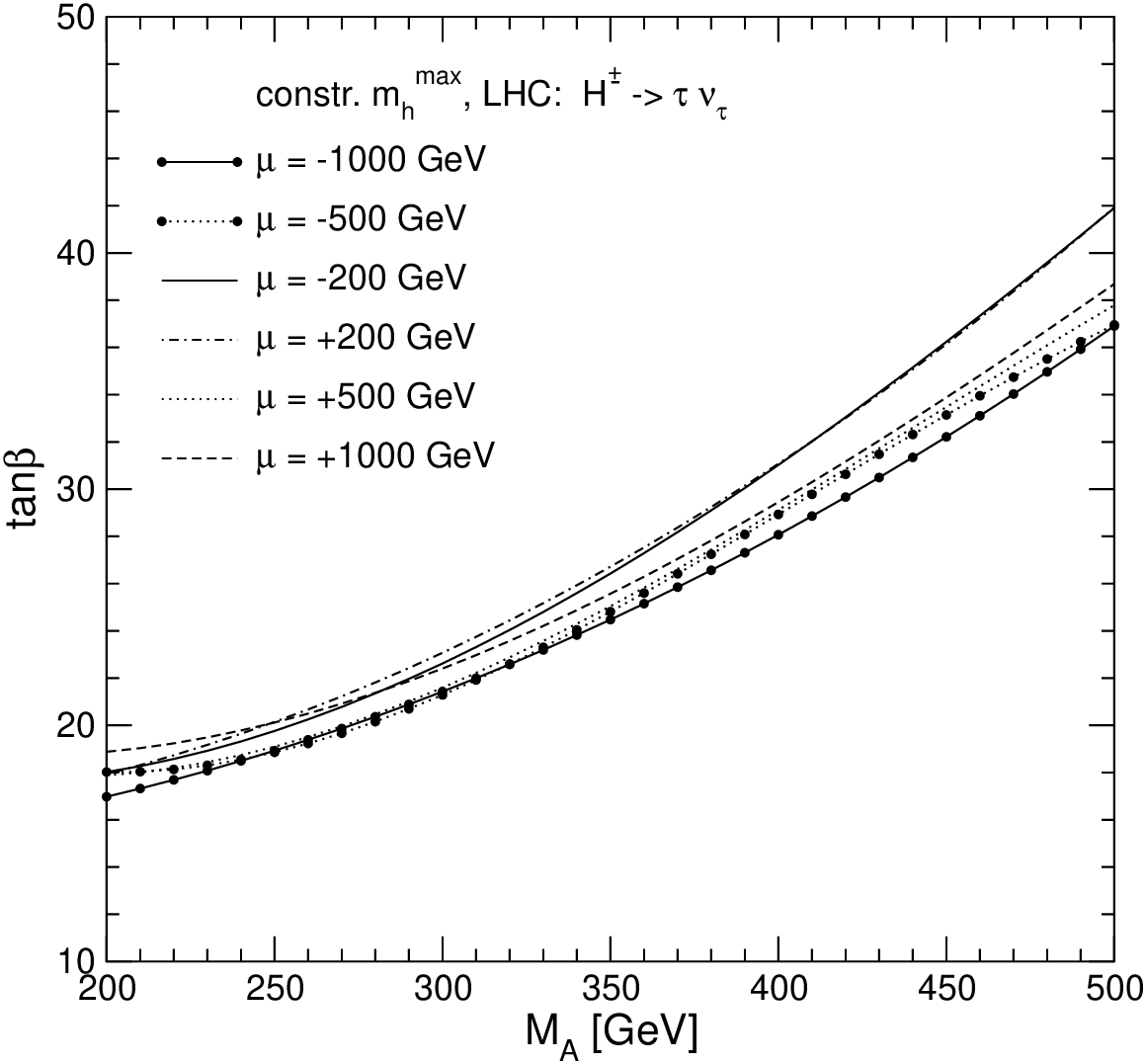}
\EC
\caption{Variation of the $5 \si$ discovery contours
for different values of $\mu$ 
in the constrained-$\mhmax$ scenario for the 
$pp \to H/A +X, \, H/A \to \tau^+\tau^-$ process (left) and 
the $H^{\pm} \to \tau \nu_\tau$ process (right).}
\label{fig:constmhmaxLHC}
\end{figure}

In \reffi{fig:constmhmaxLHC} (left) we show the results for the
constrained-$\mhmax$ scenario, see 
\refeq{constrainedmhmax}. The variation of the discovery contour in the
$\MA$--$\tb$ plane with $\mu$ is 
completely driven in this case by the
additional decay channels of the heavy Higgs bosons into charginos and
neutralinos. Correspondingly, the weakest sensitivity is obtained for the
smallest values of $|\mu|$, $\mu = \pm 200 \gev$.

\subsubsection{Discovery region for the process 
$t H^{\pm}, H^{\pm} \to \tau \nu_{\tau}$}

For this process we also refer to the analysis carried out by the CMS
Collaboration~\cite{cmshiggs,Kin2}. The corresponding analyses of the ATLAS
Collaboration can be found in \citeres{atlastdr,ketevi}. The results of
the CMS Collaboration were given for an integrated luminosity of
30~fb$^{-1}$ in the $\MA$--$\tb$ plane using the $\mhmax$ scenario with 
$\mu = -200 \gev$. No $\db$ corrections were included in the 
$gb \to t H^{\pm}$ production process~\cite{tilman} and the 
$H^{\pm} \to \tau \nu_{\tau}$ decay~\cite{hdecay}.

\begin{figure}[htb!]
\BC
\includegraphics[width=.49\textwidth]{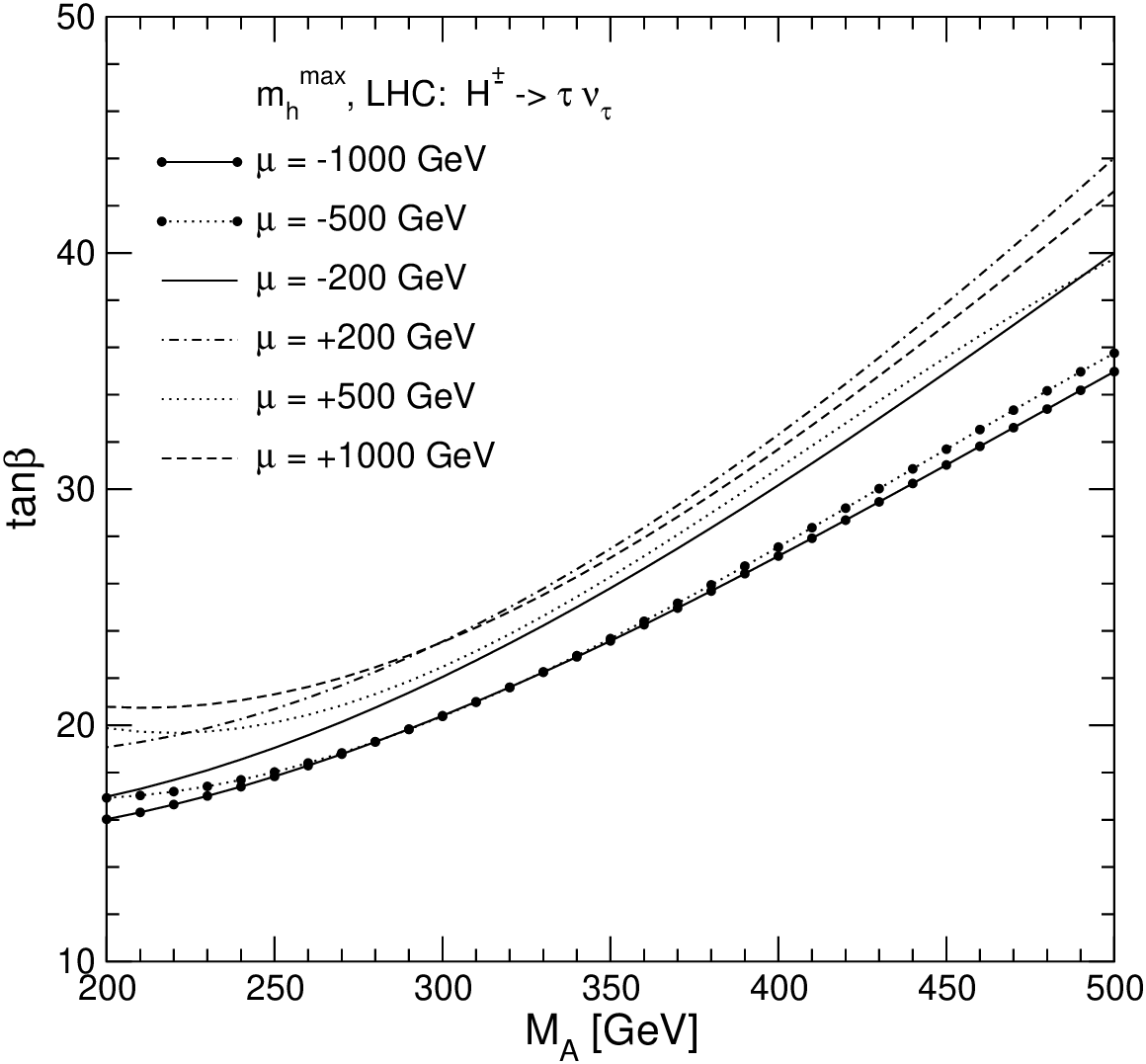}
\includegraphics[width=.49\textwidth]{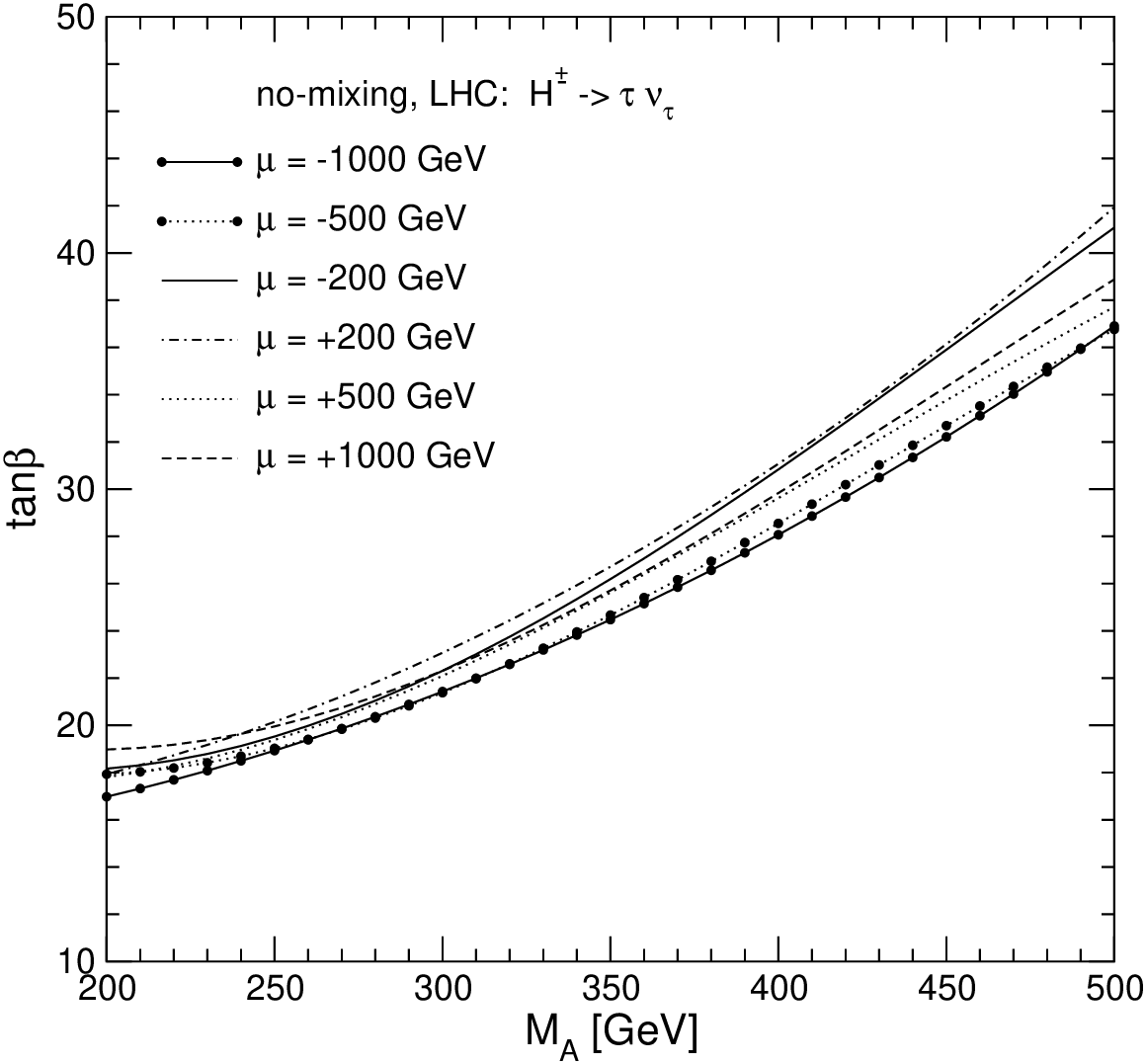}
\EC
\caption{Variation of the $5 \si$ discovery contours obtained 
from the $t H^{\pm}, H^{\pm} \to \tau \nu_{\tau}$
channel in the $\mhmax$ (left) and 
no-mixing (right) benchmark scenarios for different values of~$\mu$.}
\label{fig:LHCcharged}
\end{figure}

In \reffi{fig:LHCcharged} we investigate the impact of including the
$\db$ corrections into the production and decay processes and of varying
$\mu$.
In order to rescale the original result for the discovery reach in $\tb$ 
we have first
evaluated the $\tb$ dependence of the production and decay processes.
If no supersymmetric 
radiative corrections are included, for a fixed $\MA$ value,
the discovery potential can be inferred 
by using a rate approximately proportional to
\BE
\TQb_{\rm CMS} \times \br(H^{\pm} \to \tau \nu_{\tau})_{\rm CMS}~. 
\label{Hptaunuold}
\EE
Here 
$\tb_{\rm CMS}$ is given by the edge of the area in the $\MA$--$\tb$ plane
in which the signal $H^{\pm} \to \tau \nu_\tau$ 
is visible, as obtained in the CMS analysis. 
The $\br(H^{\pm} \to \tau \nu_{\tau})_{\rm CMS}$ has been
evaluated with {\tt HDECAY}.

The rescaled result for the discovery contour,
including all relevant $\db$ corrections, is obtained by demanding 
that the contribution
\BE
\frac{\TQb_{\rn}}{(1 + \db)^2} \times \br(H^{\pm} \to \tau \nu_{\tau})_{\rn}~,
\label{Hptaununew}
\EE
where $\db$ depends on $\tb_{\rn}$, is numerically equal to the one of
\refeq{Hptaunuold}.  The quantities in \refeq{Hptaununew} have been
evaluated with {\tt FeynHiggs}.

This procedure has been carried out in two benchmark
scenarios for various values of $\mu$. The results are shown in
\reffi{fig:LHCcharged} for the $\mhmax$~scenario (left) and for the
no-mixing scenario (right). As a consequence of the 
cancellations of the leading $\db$ effects
in the Higgs production and the Higgs decay 
the change in the discovery contour
due to the variation of
$\mu$ does not exceed $\De\tb \sim 10 (6)$ in the $\mhmax$ (no-mixing)
scenario. Also in this case there is a variation of the contour caused
by decays
into supersymmetric particles that, as in the neutral Higgs boson
case, are only relevant for small values of $|\mu|$. 
For completeness,
in \reffi{fig:AchHsusydecay} (right) we show the corresponding variation of 
the discovery contour
for the $\mhmax$ scenario as a function of $\mu$, for different
values of $\MA$.
Outside the region where the decays into supersymmetric particles are
relevant the dependence on $\mu$ is relatively weak, but somewhat more
pronounced than for the case of the neutral Higgs bosons $H$ and $A$.

In \reffi{fig:constmhmaxLHC} (right) we show the results for the
constrained-$\mhmax$ scenario, see 
\refeq{constrainedmhmax}. As in the case of the neutral Higgs bosons, the
variation of the discovery contour in the
$\MA$--$\tb$ plane is completely driven by the
additional decay channels of the heavy Higgs bosons into charginos and
neutralinos.


\section{Benchmark Scenarios}

The benchmark scenarios defined in \citere{benchmark2}, which were
mainly designed for the search for the light $\cp$-even Higgs boson $h$
in the $\cp$-conserving case,
are also useful in the search for the heavy MSSM Higgs bosons $H$, $A$
and $H^{\pm}$. 
In order to take into account the dependence on $\mu$, which as
explained above is particularly
pronounced for the $b \bar b \phi, \phi \to b \bar b$ 
channel, we suggest to extend the definition of the $\mhmax$ and
no-mixing scenarios given in \citere{benchmark2} by several discrete
values of $\mu$. The scenarios defined in \citere{benchmark2} read\\
\underline{$\mhmax:$}\\[-2em]
\BEA
\mt &=& 174.3 \gev, \non \\
\msusy &=& 1000 \gev, \non \\
\mu &=& 200 \gev, \non \\
M_2 &=& 200 \gev, \non \\
\Xt^{\OS} &=& 2\, \msusy  \; \mbox{(FD calculation)}, \non \\
\Xt^{\MS} &=& \sqrt{6}\, \msusy \; \mbox{(RG calculation)} \non \\ 
\Ab &=& \At, \non \\
\mgl &=& 0.8\,\msusy~.
\label{mhmax}
\EEA
\underline{\rm no-mixing:}\\[-2em]
\BEA
\mt &=& 174.3 \gev, \non \\
\msusy &=& 2000 \gev, \non \\
\mu &=& 200 \gev, \non \\
M_2 &=& 200 \gev, \non \\
\Xt &=& 0 \; \mbox{(FD/RG calculation)} \non \\
\Ab &=& \At, \non \\
\mgl &=& 0.8\,\msusy~.
\label{nomix}
\EEA
\underline{\rm constrained $\mhmax$:}\\[-2em]
\BEA
\mt &=& 174.3 \gev, \non \\
\msusy &=& 1000 \gev, \non \\
\mu &=& 200 \gev, \non \\
M_2 &=& 200 \gev, \non \\
\Xt^{\OS} &=& -2\, \msusy  \; \mbox{(FD calculation)}, \non \\
\Xt^{\MS} &=& -\sqrt{6}\, \msusy \; \mbox{(RG calculation)}, \non \\
\Ab &=& \At, \non \\
\mgl &=& 0.8\,\msusy~.
\label{constmhmax}
\EEA
The constrained-$\mhmax$ scenario differs from \refeq{mhmax} only by the
reversed sign of $\Xt$. 
While the positive sign of the product $(\mu\,M_2)$ results in
general in better agreement with the $(g-2)_\mu$ experimental results,
the negative sign of the product ($\mu\,\At$) yields in general 
(assuming minimal flavor violation) better agreement with the 
$\br(b \to s \ga)$ measurements.

Motivated by the analysis in \refse{sec:interpretation}
we suggest to investigate the following values of $\mu$
\BE
\mu = \pm 200, \pm 500, \pm 1000 \gev ~,
\label{eq:musuggest}
\EE
allowing both an enhancement and a suppression of the bottom Yukawa
coupling and taking into account the limits from direct searches for
charginos at LEP~\cite{pdg}.
As discussed above, the results in the constrained-$\mhmax$ scenario
are expected to yield more robust bounds against the variation of
$\mu$ than in the other scenarios.
It should be noted that the values $\mu = -500, -1000 \gev$ can lead to
such a large enhancement of the bottom Yukawa coupling that a
perturbative treatment is no longer possible in the region of very large 
values of $\tb$. Some care is therefore necessary to assess up to which
values of $\mu$ reliable results can be obtained, see e.g.\ the
discussion of \reffi{fig:cdfchange}.

The value of the top-quark mass in \citere{benchmark2} was chosen
according to the experimental central value at that time. We propose 
to substitute this value with the most up-to-date experimental
central value for $\mt$.


\section{Conclusions}

In this paper we have analyzed the impact of supersymmetric radiative
corrections on the current MSSM Higgs boson exclusion limits at the 
Tevatron and the prospective discovery reach at the LHC.
In particular, we have
studied the variation of the exclusion and discovery contours obtained 
in different MSSM benchmark scenarios under 
changes of the higgsino mass parameter
$\mu$ and the supersymmetry breaking parameters associated with the
third generation squarks. These parameters determine the most important
supersymmetric radiative corrections in the large $\tb$ region that are
associated with a change of the effective Yukawa
couplings of the bottom quarks to the Higgs fields (since the squarks
are relatively heavy in the considered benchmark scenarios, other
squark-loop effects are sub-dominant). These corrections had been
ignored or only partially considered
in some of the previous analyses of Higgs searches at hadron
colliders. We have shown that their inclusion leads to a significant
modification of the discovery and exclusion regions. 

We have investigated the exclusion bounds obtained from the Tevatron searches 
for non SM-like Higgs bosons in different channels. For the 
$b \bar b \phi, \phi \to b \bar b$ channel ($\phi = h, H, A$)
we find that the effects of
the supersymmetric radiative corrections on the exclusion bounds in the
$\MA$--$\tb$ plane are quite dramatic. While in the $\mhmax$ scenario
the current data allow to rule out values of $\tb \gsim 50$~(35) for 
$\MA \approx 100 \gev$ if the higgsino mass parameter is chosen as 
$\mu = -200 \gev$~($-1000 \gev$), hardly any bound 
on $\tb$ can be set if positive
values of $\mu$ are chosen. The shifts are smaller, but still 
important, for the no-mixing benchmark scenario.
We have shown that the 
constrained-$\mhmax$ scenario yields results that are much more stable
against variations of $\mu$ than the other benchmark scenarios.

For the inclusive channel with $\tau^+\tau^-$ final states, 
$p \bar p \to \phi \to \tau^+\tau^-$, compensations occur between large
corrections to Higgs production and decay, so that the limits in the 
$\MA$--$\tb$ plane obtained from this channel turn out to be less
affected by varying $\mu$ than the ones from the
associated production with bottom
quarks. Nevertheless we have found that the exclusion limit is shifted
by up to $\De\tb = 30$ as a consequence of choosing different input
values for $\mu$. We have investigated the impact of including the
dominant supersymmetric radiative corrections to the gluon fusion 
production process, which had previously been omitted. The inclusion of
these corrections leads to a shift of up to $\De\tb = 10$ in the 
exclusion limit. 
Following our analysis, the CDF Collaboration has adopted the
prescription outlined in this paper for incorporating the correction into
the $gg \to \phi$ production process.
The Tevatron experiments are expected to collect further data at
higher luminosities, 
up to 4--8~fb$^{-1}$, in the next few years. This will extend the Tevatron
MSSM Higgs boson discovery and exclusion reach  in the $M_A$--$\tb$ 
plane to lower values of $\tb$,  decreasing the sensitivity of the
obtained bounds to variations of the low energy supersymmetry mass parameters.

For the LHC we have analyzed the channels 
$pp \to H/A +X, \, H/A \to \tau^+\tau^-$ and 
$t H^{\pm}, H^{\pm} \to \tau \nu_{\tau}$, which yield the best
sensitivities in the search for heavy MSSM Higgs bosons. Accordingly,
the discovery contours for these channels in the $\MA$--$\tb$ plane
determine the boundary of the region where only the (SM-like) light 
$\cp$-even Higgs boson can be detected at the LHC. Since the discovery
contours for the LHC are at smaller values of $\tb$ compared to those
accessible via the
current exclusion bounds at the Tevatron, the impact of the
$\tb$-enhanced supersymmetric corrections is less pronounced
in this case. We have
studied the effect of including the dominant supersymmetric
corrections, which had been omitted in the analyses of the 
production processes at the LHC, and their variation with the relevant
parameters. Possible decays of the heavy MSSM Higgs bosons into
charginos and neutralinos have been taken into account.
We have found that the prospective discovery contours at the LHC 
are shifted by up to $\De\tb \lsim 10$.

Based on our analysis of the sensitivities of the searches for MSSM
Higgs bosons at the Tevatron and the LHC we have defined benchmark
scenarios for the analysis of MSSM Higgs-boson searches at hadron
colliders. They are based on a generalization of similar benchmark
scenarios proposed for the searches for SM-like MSSM Higgs bosons at
the Tevatron and the LHC.


\subsection*{Acknowledgements}
M.~C., C.E.M.~W.\ and G.~W.\ thank the Aspen Center for Physics for its
hospitality, where part of this work was done.
We thank
A.~Anastassov,
J.~Conway,
A.~Goussiou, 
A.~Haas,
B.~Heinemann, 
A.~Kharchilava,
R.~Kinnunen,
A.~Lath,
A.~Nikitenko,
T.~Plehn, 
M.~Schumacher 
and
M.~Spira
for helpful discussions.


\newpage
\bibliographystyle{plain}

\begin{thebibliography}{99}

\bibitem{LEPHiggsSM} [LEP Higgs working group], 
                     {\em Phys. Lett.} {\bf B 565} (2003) 61, 
                     hep-ex/0306033.

\bibitem{LEPHiggsMSSM} [LEP Higgs working group],
                   hep-ex/0107030;
                   hep-ex/0107031;
                   LHWG Note 2004-1,
                   see:\\ {\tt lephiggs.web.cern.ch/LEPHIGGS/papers} .

\bibitem{benchmark} M.~Carena, S.~Heinemeyer, C.~Wagner and G.~Weiglein,
                    hep-ph/9912223.

\bibitem{benchmark2} M.~Carena, S.~Heinemeyer, C.~Wagner and G.~Weiglein, 
                     {\em Eur. Phys. J.} {\bf C 26} (2003) 601, 
                     hep-ph/0202167.

\bibitem{tbexcl} S.~Heinemeyer, W.~Hollik and G.~Weiglein, 
                 {\em JHEP} {\bf 0006} (2000) 009,
                 hep-ph/9909540.

\bibitem{schumi} M.~Schumacher,
                 {\em Czech. J. Phys.} {\bf 54} (2004) A103.

\bibitem{D0bounds} V.~Abazov et al.  [D0 Collaboration],
                   hep-ex/0504018.

\bibitem{CDFbounds} A.~Abulencia et al.  [CDF Collaboration],
                    hep-ex/0508051.

\bibitem{Tevcharged} [CDF Collaboration],
                     hep-ex/0510065;\\
                     R.~Eusebi, Ph.d. thesis: ``Search for charged Higgs 
                     in $t \bar t$ decay products from proton-antiproton
                     collisions at $\sqrt{s}=1.96\,{\rm TeV}$'', 
                     University of Rochester, 2005.

\bibitem{Belyaev:2005nu} A.~Belyaev, J.~Pumplin, W.~Tung and C.~Yuan,
                         hep-ph/0508222.

\bibitem{stefanCM} M.~Carena, P.~Chankowski, S.~Pokorski and C.~Wagner,
                   {\em Phys. Lett.} {\bf B 441} (1998) 205, 
                   hep-ph/9805349.

\bibitem{mhiggslong} S.~Heinemeyer, W.~Hollik and G.~Weiglein,
                     {\em Eur. Phys. J.} {\bf C 9} (1999) 343,
                     hep-ph/9812472.

\bibitem{ERZ} J.~Ellis, G.~Ridolfi and F.~Zwirner,
              {\em Phys.\ Lett.} {\bf B 257} (1991) 83;\\
              Y.~Okada, M.~Yamaguchi and T.~Yanagida,
              {\em Prog.\ Theor.\ Phys. } {\bf 85} (1991) 1;\\
              H.~Haber and R.~Hempfling,
              {\em Phys.\ Rev.\ Lett.}  {\bf 66} (1991) 1815.

\bibitem{mhiggsf1lA} A.~Brignole,
                     {\em Phys. Lett.}\ {\bf B 281} (1992) 284.

\bibitem{mhiggsf1lB} P.~Chankowski, S.~Pokorski and J.~Rosiek,
                     {\em Phys. Lett.} {\bf B 286} (1992) 307;
                     {\em Nucl. Phys.} {\bf B 423} (1994) 437,
                     hep-ph/9303309.

\bibitem{mhiggsf1lC} A.~Dabelstein,
                     {\em Nucl. Phys.} {\bf B 456} (1995) 25,
                     hep-ph/9503443;
                     {\em Z. Phys.} {\bf C 67} (1995) 495,
                     hep-ph/9409375.

\bibitem{mhiggsEP1b} R.~Hempfling and A.~Hoang, 
                     {\em Phys. Lett.} {\bf B 331} (1994) 99, 
                     hep-ph/9401219.

\bibitem{mhiggsRG1a}  J.~Casas, J.~Espinosa, M.~Quir\'os and A.~Riotto,
                      {\em Nucl. Phys.} {\bf B 436} (1995) 3,
                      E: {\em ibid.} {\bf B 439} (1995) 466,
                      hep-ph/9407389.

\bibitem{mhiggsRG1} M.~Carena, J.~Espinosa, M.~Quir\'os and C.~Wagner, 
                    {\em Phys. Lett.} {\bf B 355} (1995) 209, 
                    hep-ph/9504316;\\
                    M.~Carena, M.~Quir\'os and C.~Wagner, 
                    {\em Nucl. Phys.} {\bf B 461} (1996) 407, 
                    hep-ph/9508343.

\bibitem{HHH} H.~Haber, R.~Hempfling and A.~Hoang, 
              {\em Z. Phys.} {\bf C 75} (1997) 539, 
              hep-ph/9609331.

\bibitem{mhiggsletter} S.~Heinemeyer, W.~Hollik and G.~Weiglein, 
                       {\em Phys. Rev.} {\bf D 58} (1998) 091701, 
                       hep-ph/9803277; 
                       {\em Phys. Lett.} {\bf B 440} (1998) 296, 
                       hep-ph/9807423.

\bibitem{mhiggsEP0}  R.~Zhang, 
                     {\em Phys.\ Lett. } {\bf B 447} (1999) 89, 
                     hep-ph/9808299;\\
                     J.~Espinosa and R.~Zhang, 
                     {\em JHEP} {\bf 0003} (2000) 026, 
                     hep-ph/9912236.

\bibitem{mhiggsEP1} G.~Degrassi, P.~Slavich and F.~Zwirner,
                    {\em Nucl. Phys.} {\bf B 611} (2001) 403,
                    hep-ph/0105096.

\bibitem{mhiggsEP3} J.~Espinosa and R.~Zhang,
                    {\em Nucl. Phys.} {\bf B 586} (2000) 3,
                    hep-ph/0003246. 

\bibitem{mhiggsEP2} A.~Brignole, G.~Degrassi, P.~Slavich and F.~Zwirner,
                    {\em Nucl. Phys.} {\bf B 631} (2002) 195,
                    hep-ph/0112177.

\bibitem{mhiggsEP4} A.~Brignole, G.~Degrassi, P.~Slavich and F.~Zwirner,
                    {\em Nucl. Phys.} {\bf B 643} (2002) 79,
                    hep-ph/0206101.

\bibitem{mhiggsFD2} S.~Heinemeyer, W.~Hollik, H.~Rzehak and G.~Weiglein,
                    {\em Eur. Phys. J.} {\bf C 39} (2005) 465, 
                    hep-ph/0411114;
                    hep-ph/0506254.

\bibitem{deltamb1} R.~Hempfling,
                   {\em Phys. Rev.} {\bf D 49} (1994) 6168;\\
                   L.~Hall, R.~Rattazzi and U.~Sarid,
                   {\em Phys. Rev.} {\bf D 50} (1994) 7048,
                   hep-ph/9306309;\\
                   M.~Carena, M.~Olechowski, S.~Pokorski and C.~Wagner,
                   {\em Nucl. Phys.} {\bf B 426} (1994) 269,
                   hep-ph/9402253.

\bibitem{deltamb2} M.~Carena, D.~Garcia, U.~Nierste and C.~Wagner,
                   {\em Nucl. Phys.} {\bf B 577} (2000) 577,
                   hep-ph/9912516.

\bibitem{deltamb2b} H.~Eberl, K.~Hidaka, S.~Kraml, W.~Majerotto and
                    Y.~Yamada,
                    {\em Phys. Rev.} {\bf D 62} (2000) 055006,
                    hep-ph/9912463.

\bibitem{mhiggsEP4b} G.~Degrassi, A.~Dedes and P.~Slavich,
                    {\em Nucl. Phys.} {\bf B 672} (2003) 144, 
                    hep-ph/0305127.

\bibitem{mhiggsAEC} G.~Degrassi, S.~Heinemeyer, W.~Hollik,
                    P.~Slavich and G.~Weiglein, 
                    {\em Eur. Phys. J.} {\bf C 28} (2003) 133,
                    hep-ph/0212020.

\bibitem{PomssmRep} S.~Heinemeyer, W.~Hollik and G.~Weiglein,
                    hep-ph/0412214.

\bibitem{feynhiggs} S.~Heinemeyer, W.~Hollik and G.~Weiglein,
                    {\em Comput. Phys. Comm.} {\bf 124} (2000) 76,
                    hep-ph/9812320;
                    hep-ph/0002213;
                    see: {\tt www.feynhiggs.de} .

\bibitem{feynhiggs1.2} M.~Frank, S.~Heinemeyer, W.~Hollik and 
                       G.~Weiglein,
                       hep-ph/0202166.

\bibitem{feynhiggs2} T.~Hahn, S.~Heinemeyer, W.~Hollik and G.~Weiglein,
                     hep-ph/0507009.

\bibitem{feynhiggs2.3} M.~Frank, T.~Hahn, S.~Heinemeyer, W.~Hollik,  
                       H.~Rzehak and G.~Weiglein,
                       {\em in preparation}.

\bibitem{cpsh} J.~Lee, A.~Pilaftsis et al.,
               {\em Comput. Phys. Comm.} {\bf 156} (2004) 283, 
               hep-ph/0307377.


\bibitem{bse} M.~Carena, H.~Haber, S.~Heinemeyer, W.~Hollik, C.~Wagner,
              and G.~Weiglein,
              {\em Nucl. Phys.} {\bf B 580} (2000) 29,
              hep-ph/0001002.

\bibitem{mhiggsCPXRG} A.~Pilaftsis,
                      {\em Phys. Rev.} {\bf D 58} (1998) 096010,
                      hep-ph/9803297;\\
                      A.~Pilaftsis and C.~Wagner, 
                      {\em Nucl. Phys.} {\bf B 553} (1999) 3,
                      hep-ph/9902371;\\
                      M.~Carena, J.~Ellis, A.~Pilaftsis and C.~Wagner,
                      {\em Nucl. Phys.} {\bf B 586} (2000) 92,
                      hep-ph/0003180;
                      {\em Nucl. Phys.} {\bf B 625} (2002) 345, 
                      hep-ph/0111245;\\
                      M.~Carena, J.~Ellis, S.~Mrenna, A.~Pilaftsis and 
                      C.~Wagner,
                      {\em Nucl. Phys.} {\bf B 659} (2003) 145, 
                      hep-ph/0211467;\\
                      S.~Choi, M.~Drees and J.~Lee,
                      {\em Phys. Lett.} {\bf B 481} (2000) 57,
                      hep-ph/0002287.

\bibitem{mhiggsCPXFD} S. Heinemeyer,
                      {\em Eur. Phys. J.} {\bf C 22} (2001) 521,
                      hep-ph/0108059;\\
                      M.~Frank, S.~Heinemeyer, W.~Hollik and G.~Weiglein,
                      hep-ph/0212037,
                      appeared in the proceedings of SUSY02,
                      DESY, Hamburg, Germany, July 2002.

\bibitem{habilSH} S.~Heinemeyer,
                  hep-ph/0407244.

\bibitem{mhiggsEP5} S.~Martin, 
                    {\em Phys. Rev.} {\bf D 65} (2002) 116003,
                    hep-ph/0111209;
                    {\em Phys. Rev.} {\bf D 66} (2002) 096001,
                    hep-ph/0206136;
                    Phys. Rev. {\bf D 67} (2003) 095012, 
                    hep-ph/0211366;
                    {\em Phys. Rev.} {\bf D 68} (2003) 075002, 
                    hep-ph/0307101; 
                    {\em Phys. Rev.} {\bf D 70} (2004) 016005, 
                    hep-ph/0312092;
                    {\em Phys. Rev.} {\bf D 71} (2005) 016012, 
                    hep-ph/0405022;
                    {\em Phys. Rev.} {\bf D 71} (2005) 116004, 
                    hep-ph/0502168;\\
                    S.~Martin and D.~Robertson,
                    hep-ph/0501132.

\bibitem{mhiggslle} S.~Heinemeyer, W.~Hollik and G.~Weiglein,
                    {\em Phys. Lett.} {\bf B 455} (1999) 179,
                    hep-ph/9903404.

\bibitem{deltamb3} J.~Guasch, P.~H\"afliger and M.~Spira, 
                   {\em Phys.\ Rev.} {\bf D 68} (2003) 115001,
                   hep-ph/0305101.

\bibitem{ggHSM} S.~Dawson and S.~Willenbrock,
                {\em Phys. Rev.} {\bf D 40} (1989) 2880;\\
                S.~Dawson,
                {\em Nucl. Phys.} {\bf B 359} (1991) 283;\\
                A.~Djouadi, M.~Spira and P.~Zerwas,
                {\em Phys. Lett.} {\bf B 264} (1991) 440;\\
                S.~Catani, D.~de Florian and M.~Grazzini,
                {\em JHEP} {\bf 0105} (2001) 025, 
                hep-ph/0102227;\\
                R.~Harlander and W.~Kilgore,
                {\em Phys. Rev. Lett.} {\bf 88} (2002) 201801, 
                hep-ph/0201206;\\
                C.~Anastasiou and K.~Melnikov,
                {\em Nucl. Phys.} {\bf B 646} (2002) 220, 
                hep-ph/0207004;\\
                V.~Ravindran, J.~Smith and W.~van Neerven,
                {\em Nucl. Phys.} {\bf B 665} (2003) 325, 
                hep-ph/0302135.

\bibitem{ggHMSSM} R.~Harlander and M.~Steinhauser,
                  {\em Phys. Lett.} {\bf B 574} (2003) 258, 
                  hep-ph/0307346;
                  {\em JHEP} {\bf 0409} (2004) 066, 
                  hep-ph/0409010.

\bibitem{D0bbhSM} J.~Campbell, R.~Ellis, F.~Maltoni and S.~Willenbrock,
                  {\em Phys.\ Rev.}  {\bf D 67} (2003) 095002,
                  hep-ph/0204093.

\bibitem{bbHSM} S.~Dittmaier, M.~Kr\"amer and M.~Spira,
                {\em Phys. Rev.} {\bf D 70} (2004) 074010, 
                hep-ph/0309204;\\
                S.~Dawson, C.~Jackson, L.~Reina and D.~Wackeroth,
                {\em Phys. Rev.} {\bf D 69} (2004) 074027, 
                hep-ph/0311067;
                {\em Phys. Rev. Lett.} {\bf 94} (2005) 031802, 
                hep-ph/0408077;\\
                J.~Campbell et al.,
                hep-ph/0405302.

\bibitem{robikilgore} R.~Harlander and W.~Kilgore,
                      {\em Phys. Rev.} {\bf D 68} (2003) 013001, 
                      hep-ph/0304035.

\bibitem{hff} S.~Gorishny, A.~Kataev, S.~Larin and L.~Surguladze,
              {\em Mod. Phys. Lett.} {\bf A 5} (1990) 2703;
              {\em Phys. Rev.} {\bf D 43} (1991) 1633;\\
              A.~Kataev and V.~Kim, 
              {\em Mod. Phys. Lett.} {\bf A 9} (1994) 1309;\\
              L.~Surguladze,
              {\em Phys. Lett.} {\bf B 338} (1994) 229,
              hep-ph/9406294;
              {\em Phys. Lett.} {\bf B 341} (1994) 60,
              hep-ph/9405325;\\
              K.~Chetyrkin, 
              {\em Phys. Lett.} {\bf B 390} (1997) 309,
              hep-ph/9608318;\\
              K.~Chetyrkin and A.~Kwiatkowski,
              {\em Nucl. Phys.} {\bf B 461} (1996) 3,
              hep-ph/9505358;\\
              S.~Larin, T.~van~Ritbergen and J.~Vermaseren,
              {\em Phys. Lett.} {\bf B 362} (1995) 134,
              hep-ph/9506465;\\
              P.~Chankowski, S. Pokorski and J. Rosiek, 
              {\em Nucl. Phys.} {\bf B 423} (1994) 497;\\
              S.~Heinemeyer, W.~Hollik and G.~Weiglein, 
              {\em Eur. Phys. J.} {\bf C 16} (2000) 139, 
              hep-ph/0003022.
	      
\bibitem{deltamb4} M.~Carena, S.~Mrenna and C.~Wagner,
                   {\em Phys. Rev.} {\bf D 60} (1999) 075010,
                   hep-ph/9808312;
                   {\em Phys. Rev.} {\bf D~62} (2000) 055008,
                   hep-ph/9907422.

\bibitem{SolaChargedHiggs} A.~Belyaev, D.~Garcia, J.~Guasch and J.~Sola,
                           {\em JHEP} {\bf 0206} (2002) 059, 
                           hep-ph/0203031.

\bibitem{Belyaev:2005ct} A.~Belyaev, A.~Blum, R.~Chivukula and E.~Simmons,
                         hep-ph/0506086.

\bibitem{higlu} M.~Spira,
                hep-ph/9510347.

\bibitem{cdfold} [CDF Collaboration],
                 CDF note 7161.

\bibitem{atlastdr} ATLAS Collaboration,
        {\em Detector and Physics Performance Technical Design Report},
        CERN/LHCC/99-15 (1999), see:\\
        {\tt atlasinfo.cern.ch/Atlas/GROUPS/PHYSICS/TDR/access.html}~.

\bibitem{cmshiggs} S.~Abdullin et al.,
                   {\em Eur. Phys. J.} {\bf C 39S2} (2005) 41.

\bibitem{filip} A.~Datta, A.~Djouadi, M.~Guchait and F.~Moortgat,
                {\em Nucl. Phys.} {\bf B 681} (2004) 31, 
                hep-ph/0303095.

\bibitem{filip2} F.~Moortgat {\em in}
                 G.~Weiglein et al. [LHC / LC Study Group],
                 hep-ph/0410364.

\bibitem{KinNik} R.~Kinnunen and A.~Nikitenko,
                 CMS note 2003/006.

\bibitem{atlasnote} J.~Thomas, 
                    ATL-PHYS-2003-003;\\
                    D.~Cavalli and D.~Negri, 
                    ATL-PHYS-2003-009.

\bibitem{hdecay} A.~Djouadi, J.~Kalinowski and M.~Spira,
                 {\em Comput. Phys. Comm.} {\bf 108} (1998) 56,
                 hep-ph/9704448.

\bibitem{hqq} See: {\tt people.web.psi.ch/spira/hqq}~.

\bibitem{Kin2} R.~Kinnunen, P.~Salmi and N.~Stepanov,
               CMS note 2002/024.

\bibitem{ketevi} K.~Assamagan, Y.~Coadou and A.~Deandrea,
                 {\em Eur.\ Phys.\ J.\ directC} {\bf 4} (2002) 9,
                 hep-ph/0203121;\\
                 K.~Assamagan and N.~Gollub,
                 {\em Eur.\ Phys.\ J.} {\bf C 39S2} (2005) 25,
                 hep-ph/0406013.

\bibitem{tilman} R.~Kinnunen, S.~Nikitenko and T.~Plehn,
                 private communication.

\bibitem{pdg} G.~Abbiendi et al.\ [OPAL Collaboration],
              {\em Eur. Phys. J.} {\bf C 35} (2004) 1, 
              hep-ex/0401026.

\end{thebibliography}


\end{document}